\documentclass[prb,preprint,eqsecnum]{revtex4-1} 
\usepackage{graphicx} 
\usepackage{epstopdf}
\newcommand{\be}{\begin{equation}}
\newcommand{\ee}{\end{equation}}
\newcommand{\ba}{\begin{eqnarray}}
\newcommand{\ea}{\end{eqnarray}}
\newcommand{\nn}{\nonumber}
\newcommand{\la}{\label} 
\newcommand{\e}{{\rm e}}
\newcommand{\ac}{{\bf a}}
\newcommand{\om}{{\omega}}
\newcommand{\bp}{{\bf p}}
\newcommand{\bq}{{\bf q}}

\newcommand{\bv}{{\bf v}}

\newcommand{\bac}{{\bf a}}

\newcommand{\bF}{{\bf F}}

\newcommand{\br}{{\bf r}}

\newcommand{\dt}{{\Delta t}}
\newcommand{\ep}{{\epsilon}}
\newcommand{\ct}{{\cal T}}
\newcommand{\dl}{{\delta}}
\newcommand{\hH}{{\hat H}}
\newcommand{\hT}{{\hat T}}
\newcommand{\cT}{{\cal T}}
\newcommand{\hV}{{\hat V}}

\newcommand{\tH}{{\hat H}}

\newcommand{\pa}{{\partial}}

\newcommand{\lra}{\longrightarrow}

\begin{document}

\title{Structure of numerical algorithms and advanced mechanics}

\author{Siu A. Chin}
\affiliation{Department of Physics and Astronomy, Texas A\&M University,
College Station, TX 77843, USA}

\email{chin@physics.tamu.edu} 

\begin{abstract}

Most elementary numerical schemes found useful for solving classical trajectory problems 
are {\it canonical transformations}. This fact should be make more widely known among
teachers of computational physics and Hamiltonian mechanics.
From the perspective of advanced mechanics, there are no bewildering number of seemingly 
arbitrary elementary schemes based on Taylor's expansion. There are only {\it two} canonical 
first and second order algorithms, on the basis of which one can comprehend the structures  
of higher order symplectic and non-symplectic schemes. This work shows that, from
the most elementary first-order methods to the most advanced fourth-order algorithms,
all can be derived from canonical transformations and Poisson brackets of advanced mechanics.

\end{abstract}
\maketitle

\section {Introduction}
\la{int}

More than 30 years ago, there was a flurry of discussion in this journal
on the merits of various elementary numerical schemes for solving
particle trajectories in classical mechanics.\cite{cro81,lar83,bel83,sta84} 
The focus was on solving Newton's equation-of-motion 
\begin{equation}
m\ddot{x_i}=F(x_i),
\label{newton}
\end{equation}
as a second-order differential equation.
The effectiveness of these numerical schemes 
were not well understood in that seemingly minor adjustments in the algorithm,
without altering the order of the Taylor expansion, can drastically change the stability of the
algorithm. The history of each numerical scheme, especially that of Cromer, \cite{cro81}
seems to suggest that good algorithms are just a matter of accidental 
discovery or personal insight. To many readers of this journal,
 there does not seem to be order and structure to the study of 
numerical methods, only endless Taylor expansions with arbitrary term arrangements.
  
This impression is incorrect. Without exception, all elementary algorithms
up to the second order that have been found useful in solving Newton's equation-of-motion, such
as Cromer \cite{cro81}, velocity-Verlet\cite{gou07}, leap-frog, etc., 
are all {\it canonical transformations} (CT), or 
in modern terms, {\it symplectic integrators} (SI). All can be derived systematically 
from a single operator solution to Poisson's equation-of-motion. 
There is no arbitrariness at all. There are only two standard, literally ``canonical", 
first and second-order algorithms, out of which nearly all symplectic
and non-symplectic Runge-Kutta-Nystr\"om (RKN) algorithms 
can be derived.  While this ``modern synthesis" 
has been known for some time in the literature, it has yet to impact the teaching of 
computational physics at the secondary or undergraduate level. The purpose of this article 
is to make it clear, that even the most elementary numerical algorithms, are deeply rooted 
in advanced mechanics. Knowing this synthesis will therefore enrich {\it both} the 
teaching of computational physics\cite{tim08} and advanced mechanics. \cite{jor04}

The topic of symplectic integrators has been reviewed in this journal by 
Donnelly and Rogers.\cite{don05} They begin their discussion
with the symplectic form of Hamilton's equations and give the impression
that symplectic integrators are mostly a modern development from the 1990's. 
This work will provide a more measured
introduction, emphasizing the fact that first-order algorithms
are as old as canonical transformations themselves. Also, both
Donnelly and Rogers\cite{don05} and Timberlake and Hasbun\cite{tim08}
begin with symplectic integrators and explain their stability as due to  
Liouville's theorem \cite{gol80,lan60} (to be defined in the next section).
 While this explanation is correct, the stability of
any numerical scheme can be best understood in terms of the algorithm's 
error propagation. This work shows that if stability is required, with no growth
in propagation errors, then the algorithm must obey Liouville's theorem.
The simplest way in which one can ensure Liouville's theorem is to
require algorithms to be canonical transformations. Thus this work arrives at symplectic 
integrators from the requirement of numerical stability, rather than 
postulating symplectic schemes then explaining their stabilities.

In order to show the relevance of advanced mechanics to the structure of algorithms,
this work also seeks to demystify the study of Poissonian mechanics. Poisson brackets 
have been tucked away in back pages of graduate texts, seemingly for the sole
purpose of demonstrating their similarity to quantum commutator brackets. 
This impression is also incorrect. Poissonian mechanics provides the simplest way of 
generating canonical/symplectic transformations and is therefore the most practical 
mean of deriving stable numerical algorithms. This work aims to show how
simple Poisson brackets really are, if one just do away with the bracket notation.

Finally, this work seeks to bring the readers up-to-date, on the most advanced 
numerical algorithms currently available for solving classical dynamics.

Beginning 
in Section \ref{stab}, we discuss how errors propagate in any numerical scheme
and show that stability requires that the determinant of the algorithm's Jacobian matrix be one,
which is Liouville's theorem.\cite{gol80,lan60} Furthermore, this requirement is
easily satisfied if the dynamic variables, positions and momenta, are updated {\it sequentially}. 
We discuss canonical transformations in Section \ref{can} and show that one of the
hallmark of CT (overlooked in all textbooks) is that they are {\it sequential}. The two
fundamental first-order algorithms are then derived. 
The use of Poisson brackets is introduced in Section \ref{lie} and second and higher order SI
are derived in Sections \ref{trh} and \ref{hisym} on the basis of {\it time-reversibility}.
In Section \ref{rkn}, we show that
non-symplectic RKN schemes can also be derived using the same 
operator formalism as SI. In Section \ref{ftal}, the most recent development
of {\it forward time-step} SI is outlined.  
Section \ref{sum} then summarizes key findings of this work.

\section {Stability of numerical algorithms}
\la{stab}

Let's start with a one-dimensional problem with only a spatial-dependent potential $V(x)$.
In this case, Newton's second law reads
\begin{equation}
m\ddot{x}=F(x)=-\frac{\partial V}{\partial x}.
\label{newton2}
\end{equation}
Through out this work, we will let $\dot x$ and $\ddot x$ denote the {\it kinematic 
velocity} and {\it acceleration} and reserve the symbol {\it v} and {\it a} for denoting
$$
v\equiv\frac{p}{m}\qquad{\rm and}\qquad a(x)\equiv\frac{F(x)}{m}.
$$
From (\ref{newton2}), Newtonian dynamics is just a matter of solving a
second-order differential equation:
$$
\ddot x=a(x),
$$
which can be decoupled into a pair of equations
$$
 \dot x=v \qquad{\rm and}\qquad \dot v=a(x).
$$
If $x_n$ and $v_n$ are the values at $t=n\dt$, then 
a truncation of Taylor's expansion to first order in $\dt$, 
$$x(t+\dt)=x(t)+\dot x(t)\dt+O(\dt^2)$$ 
yields the Euler algorithm
\be
x_{n+1}=x_n+v_n\dt,\qquad v_{n+1}=v_n+a_n\dt,
\la{euler}
\ee
where $a_n=a(x_n)$. If $a_n$ is not constant, this algorithm is {\it unstable} for any $\dt$, 
no matter how small. Let's see why. Let $x_n=\tilde x_n+\ep_n$ 
and $v_n=\tilde v_n+\dl_n$, where $\tilde x_n$ are $\tilde v_n$ are 
the exact values at time step $n$, and $\ep_n$ and $\dl_n$ are the errors 
of $x_n$ and $v_n$ produced by the algorithm.
For a general one-step algorithm, 
$$
x_{n+1}=x_{n+1}(x_n,v_n),\qquad
v_{n+1}=v_{n+1}(x_n,v_n),
$$
and therefore
\ba 
\tilde x_{n+1}+\ep_{n+1}&=&x_{n+1}(x_n+\ep_n,v_n+\dl_n)\nn\\
&=&x_{n+1}(x_n,v_n)+\frac{\partial x_{n+1}(x_n,v_n)}{\partial x_n}\ep_n
                   +\frac{\partial x_{n+1}(x_n,v_n)}{\partial v_n}\dl_n+\cdots .
\nn
\ea
If the numerical scheme is exact, then $x_{n+1}(x_n,v_n)=\tilde x_{n+1}$. If not,
the difference is the {\it truncation error} of the algorithm. For a first-order algorithm,
the truncation error would be $O(\dt^2)$. Let's ignore this {\it systematic} error of the
algorithm and concentrate on how the errors propagate from one step to the next.
Doing the same analysis for $v_{n+1}$ then yields
$$
\left(\begin{array}{c}
       \ep_{n+1}\\
       \dl_{n+1}
      \end{array}\right)
=M
\left(\begin{array}{c}
       \ep_n\\
       \dl_n
      \end{array}\right),
$$
where the {\it errors propagation} matrix
$$
M=\left(\begin{array}{cc}
                \frac{\partial x_{n+1}}{\partial x_n}  & \frac{\partial x_{n+1}}{\partial v_n} \\
               \frac{\partial v_{n+1}}{\partial x_n}   & \frac{\partial v_{n+1}}{\partial v_n} 
      \end{array}\right)\equiv {{\partial(x^{n+1},v^{n+1})}\over{\partial(x^n,v^n)} },
$$
is identical to the {\it Jacobian} matrix for the transformation from $(x_n,v_n)$ to $(x_{n+1},v_{n+1})$, that is,
\be
dx_{n+1}dv_{n+1}={\rm det} M\, dx_{n}dv_{n}.
\la{jacm}
\ee

For the harmonic oscillator, the transformation $(x_n,v_n)\rightarrow(x_{n+1},v_{n+1})$
is linear in $\{x_n, v_n\}$ and $M$ is a constant matrix. Hence,
\ba
\left(\begin{array}{c}
       \ep_{n}\\
       \dl_{n}
      \end{array}\right)
      &=& M^n
\left(\begin{array}{c}
       \ep_0\\
       \dl_0
      \end{array}\right)
      = \left[S^{-1}\left(\begin{array}{cc}
                \lambda_1  & 0 \\
                  0   & \lambda_2 
      \end{array}\right)S\right]^n
\left(\begin{array}{c}
       \ep_0\\
       \dl_0
      \end{array}\right),
\nn\\
     &=& S^{-1}\left(\begin{array}{cc}
                \lambda_1^n  & 0 \\
                  0   & \lambda_2^n 
      \end{array}\right)S
\left(\begin{array}{c}
       \ep_0\\
       \dl_0
      \end{array}\right),
\nn
\ea
where $M$ has been diagonalized with eigenvalues $\lambda_1$ and $\lambda_2$.
Thus the propagation errors of the algorithm grow exponentially with the magnitude of the
eigenvalues. Since $M$ is a $2\times 2$ matrix, the eigenvalues are solutions to
the quadratic equation given by
\be
\lambda_{1,2}=\frac{T}2 \pm\sqrt{\left(\frac{T}2\right)^2 -D},
\la{lam}
\ee
where $T$ and $D$ are the trace and determinant of $M$ respectively.
When the eigenvalues are complex, they must be complex
conjugates of each other, $\lambda_2=\lambda_1^*$, with the same modulus
$\lambda=\sqrt{\lambda_1\lambda_2}$. But $\lambda_1\lambda_2=\det M$, therefore,
the error will {\it not} grow, {\it i.e.}, the algorthm will be stable, {\it only if}
its eigenvalues have unit modulus, or $\det M=1$. In this case, one can rewrite (\ref{lam}) as 
$$
\lambda_{1,2}=\frac{T}2 \pm i\sqrt{1-\left(\frac{T}2\right)^2}
=\cos\theta \pm i\sin\theta=\e^{\pm i \theta},
$$
where one has defined $\cos\theta\equiv T/2$, which is possible whenever $|T/2|\le 1$.
Thus when the eigenvalues are complex, one can conclude that:
1) the algorithm can be stable only if $\det M=1$, and 2)
the range of stability is given by $|T/2|\le 1$. Outside of this range of stability, when $|T/2|>1$, 
the eigenvalues are real given by (\ref{lam}), with one modulus
greater than one. 

Let's examine the stability of Euler's algorithm (\ref{euler}) in this light. 

For the harmonic oscillator, $a_n=-\omega^2x_n$, where $\omega^2=k/m$. 
The Jacobian, or the error-propagation matrix, for the Euler algorithm is
$$
M=\left(\begin{array}{cc}
                1  & \dt \\
               -\omega^2\dt  & 1 
      \end{array}\right),
$$
with $\det M=1+\omega^2\dt^2$ and $|\lambda_{1,2}|=\sqrt{1+\omega^2\dt^2}>1$.
Thus the algorithm is {\it unstable} at any finite $\dt$, no matter how small.

By constrast, Cromer's algorithm\cite{cro81} corresponding to 
\be
v_{n+1}=v_n+a_n\dt,\qquad  x_{n+1}=x_n+v_{n+1}\dt,
\la{cromer}
\ee
when applied to the harmonic oscillator, has the Jacobian matrix
$$
M=\left(\begin{array}{cc}
                1-\omega^2\dt^2  & \dt \\
               -\omega^2\dt  & 1 
      \end{array}\right)
$$
with $\det M=1$.
It differs from the Euler algorithm only in using the updated $v_{n+1}$ to
compute the new $x_{n+1}$. 
Since here $T/2=1-\omega^2\dt^2/2$, when $\dt$ increases
from zero, the algorithm is stable as the two eigenvalues $\e^{\pm i\theta}$ 
start out both equal to 1 at $\theta=0$ and race above and below the unit complex
circle until both are equal to -1 at $\theta=\pi$. This point defines the largest
$\dt$ for which the algorithm is stable: $1-\omega^2\dt^2/2=-1$,
or $\dt=2/\omega$, which is nearly a third of the harmonic oscillator period $P=2\pi/\omega$. 
For $\dt$ greater than this, $T/2<-1$, the eigenvalue 
$T/2 -\sqrt{(T/2)^2 -1}$ from (\ref{lam}) will have modulus greater than one
and the algorithm is then unstable. (In this work, stability will always mean
{\it conditional} stability, with $\dt$ sufficiently small such that $|T/2|<1$.)

The condition det$M=1$ implies that in (\ref{jacm}) $dx_n dv_n=dx_{n+1}dv_{n+1}$. 
This means that the {\it phase-space} formed by the dynamical variables $x$ and $v=p/m$ is
unaltered, not expanded nor contracted by the algorithm. This is an exact property of
Hamiltonian mechanics,\cite{gol80,lan60} known as {\it Liouville's Theorem}.
We have therefore arrived at an unexpected connection, that {\it a stable numerical algorithm for
solving classical dynamics must preserve the phase-space volume with {\rm det}$M=1$}. 

In the general case where $M$ is no longer a constant, Cromer's algorithm
maintains $\det M=1$ for an arbitrary $a(x)$ and is always stable at a sufficiently
small $\dt$. 

How then should one devise algorithms with det$M=1$? This is clearly not guaranteed by
just doing Taylor expansions, as in Euler's algorithm. Let's see how Cromer's algorithm 
accomplishes this.
 
The distinctive characteristic of Cromer's algorithm is that it updates each of the dynamical
variables {\it sequentially}. Surprisingly, this is sufficient to guarantee  det$M=1$.
By updating sequentially, (\ref{cromer}) can be viewed as a sequence
of {\it two} transformations. The first, updating only $v_n$, is the transformation
$(x_n,v_n)\rightarrow(x^*,v^*)$
$$
v^*=v_n+a(x_n)\dt,\quad  x^*=x_n\quad{\rm with}\quad  
M_1=\left(\begin{array}{cc}
                1  & 0 \\
               a'(x_n)\dt  & 1 
      \end{array}\right).
$$
The second, updating only $x_n$, is the transformation 
$(x^*,v^*)\rightarrow(x_{n+1},v_{n+1})$
$$
v_{n+1}=v^*,\quad  x_{n+1}=x^*+v^*\dt \quad{\rm with}\quad  
M_2=\left(\begin{array}{cc}
                1  & \dt \\
               0  & 1 
      \end{array}\right).
$$
Since each Jacobian matrix obviously has $\det M_{1,2}=1$, their product, corresponding to
the Jacobian matrix of the two transformations, must have $\det M=\det M_1\det M_2=1$.
In the harmonic oscillator example, we have
indeed
$$
\left(\begin{array}{cc}
                1  & \dt \\
               0  & 1 
      \end{array}\right)
\left(\begin{array}{cc}
                1  & 0 \\
               -\omega^2\dt  & 1 
      \end{array}\right)=
      \left(\begin{array}{cc}
                1-\omega^2\dt^2  & \dt \\
               -\omega^2\dt  & 1 
      \end{array}\right).
$$

When there are $n$ degrees of freedom with ${\bf a}({\bf x})$, one then has the generalization
\be
M_1=\left(\begin{array}{cc}
                {\bf I}  & {\bf 0} \\
                \frac{\partial a_i({\bf x}_n)}{\partial x_j}\dt  & {\bf I} 
      \end{array}\right)
 \qquad     
 M_2=\left(\begin{array}{cc}
                {\bf I}  & \dt {\bf I} \\
                {\bf 0} & {\bf I} 
      \end{array}\right),
\la{nmat}
\ee
where {\bf I} is the $n\times n$ unit matrix. Again one has det$M_{1,2}=1$ and 
$\det M=\det M_1\det M_2=1$.

Summarizing our finding thus far: Stability requires $\det M=1$, and
$\det M=1$ is guaranteed by sequential updating with determinants (\ref{nmat}). 
The instability of Euler's algorithm is precisely that its updating is simultaneous and {\it not} sequential.
We will see in the next section that sequential updating is a hallmark of 
canonical transformations, and for the usual separable Hamiltonian (\ref{ham}) below, the 
determinants of CT are of the form (\ref{nmat}).

\section {Canonical transformations and stability}
\la{can}

Hamiltonian mechanics is the natural arena for developing
stable numerical algorithm because of its key idea of canonical transformations.
To get to Hamiltonian mechanics, we must first go through the 
Lagrangian formulation, where Cartesian particle positions, such as $x_i$,
are first broaden to {\it generalized coordinates} $q_i$, such as 
angles, to automatically satisfy constraints. The trajectory is then determined by
the Euler-Lagrangian equation \cite{gol80,lan60}
\begin{equation}
\dot{p_i}={{\partial L}\over{\partial q_i}},
\label{langeq}
\end{equation}
where $p_i$ is the {\it generalized momentum} defined by
\begin{equation}
p_i={{\partial L}\over{\partial \dot q_i}},
\label{genp}
\end{equation}
and where $L$ is the Lagrangian function 
$$
L={1\over 2}m\sum_{i=1}^n\dot q_i^2-V(q_i).
$$
One can easily check that, when $q_i$ are just Cartesian coordinates $x_i$, (\ref{langeq})
reproduces Newton's second law (\ref{newton}). 

In Hamiltonian formulation of mechanics, the generalized 
momentum $p_i$  as defined by (\ref{genp}), is elevated as a separate and {\it independent} degree of 
freedom, no longer enslaved to the generalized coordinate as proportional to its
time-derivative.	 This entails that one
replaces all $\dot q_i$ by $p_i$ in the Hamiltonian function
\begin{eqnarray}
H(p_i,q_i)&&=\sum_{i=1}^n p_i\dot q_i-L(q_i,\dot q_i),\nonumber\\
      &&={1\over{2m}}\sum_{i=1}^n p_i^2+V(q_i),
\label{ham}
\end{eqnarray}
from which classical dynamics is evolved by Hamilton's equations,
\begin{equation}
\dot q_i={{\partial H}\over{\partial p_i}},\qquad    
\dot p_i=-{{\partial H}\over{\partial q_i}}.
\label{hameq}
\end{equation}
Again, if $q_i$ are just Cartesian coordinates, then the above reduces back to
Newton's equation of motion (\ref{newton}). 
It is of paramount importance that $q_i$ and $p_i$ are now
treated as equally fundamental, and independent, dynamical variables. 

Just as some geometric problems are easier to solve by transforming from Cartesian coordinates
to spherical coordinates, some dynamical
problems are easier to solve by transformations that completely mix up
the $n$ generalize momenta $p_i$ and the $n$ generalize coordinates $q_i$. 
In order for the transformed variables to solve the same dynamical problem, 
they must obey Hamilton's equation with respect to the Hamiltonian function of the
transformed variables. Transformations that can do this are 
called {\it canonical}. Canonical transformations make full use of the
$2n$ degrees-of-freedom of Hamiltonian dynamics and are the fundamental building
blocks of stable numerical schemes.

For the standard Hamiltonian (\ref{ham}),
a canonical transformation $(q_i,p_i)\rightarrow (Q_i,P_i)$ is a transformation such that
Hamilton's equations are preserved for the new variables
$$
\dot Q_i={{\partial K}\over{\partial P_i}},\qquad    
\dot P_i=-{{\partial K}\over{\partial Q_i}},
$$
with respect to the transformed Hamiltonian $K(Q_i,P_i)=H(q_i,p_i)$. 

Historically, canonical transformations were first formulated in 
in terms of four types of {\it generating} functions\cite{gol80, lan60},
$F_1(q_k,Q_k,t)$, $F_2(q_k,P_k,t)$, $F_3(p_k,Q_k,t)$, $F_4(p_k,P_k,t)$. 
It would take us too afar field to derived these transformation equations from
first principle, so we will simply state them. For our purpose, it is only necessary to 
consider canonical transformations generated by $F_2$ and $F_3$ without any explicit time-dependence.
The transformation equations are then
\be
p_i=\frac{\partial F_2(q_k,P_k)}{\partial q_i},\qquad 
Q_i=\frac{\partial F_2(q_k,P_k)}{\partial P_i},
\la{f2}
\ee
and
\be
q_i=-\frac{\partial F_3(p_k,Q_k)}{\partial p_i},\qquad 
P_i=-\frac{\partial F_3(p_k,Q_k)}{\partial Q_i}.
\la{f3}
\ee
Consider first the case of $F_2$.
The first equation in (\ref{f2}) is an implicit
equation for determining $P_i$ in terms of $q_i$ and $p_i$. 
This can be considered as a transformation $(q_i, p_i)\rightarrow (q^*_i, p^*_i)$
with $q^*_i=q_i$ and $p^*_i=P_i$.
The second equation in (\ref{f2}) is an explicit equation for determining $Q_i$ in 
terms of $q^*_i$ and $p^*_i$. This can be considered as a transformation 
$(q^*_i, p^*_i)\rightarrow (Q_i,P_i)$.
Their respective Jacobian matrices are therefore
\be
M_1=\left(\begin{array}{cc}
                {\bf I}  & {\bf 0} \\
                \frac{\partial P_i}{\partial q_j} & \frac{\partial P_i}{\partial p_j}
      \end{array}\right),
      \qquad
 M_2=\left(\begin{array}{cc}
                \frac{\partial Q_i}{\partial q_j} & \frac{\partial Q_i}{\partial P_j} \\
                 {\bf 0} & {\bf I}
      \end{array}\right),
\la{ctmat}
\ee
with determinant
\ba
\det M=\det M_2\det M_1&=&
\det \left( \frac{\partial Q_i}{\partial q_j}\right)
\det \left(\frac{\partial P_i}{\partial p_j}\right)\nn\\
&=&
\det \left( \frac{\partial Q_i}{\partial q_j}\right)/
\det \left(\frac{\partial p_j}{\partial P_i}\right)=1.\nn
\ea
The last line follows because:  1)  $(\partial p_j/\partial P_i)$ is the inverse matrix of
$(\partial P_i/\partial p_j)$  
$$\sum_{i=1}^n\frac{\partial p_j}{\partial P_i}\frac{\partial P_i}{\partial p_k}=\frac{\partial p_j}{\partial p_k}=\delta_{jk},
$$ 
and 2)
$$\frac{\partial Q_i}{\partial q_j}=\frac{\partial^2 F_2}{\partial q_j\partial P_i}
=\frac{\partial p_j}{\partial P_i}.
$$
This sequential way of guaranteeing $\det M=1$ for a {\it general} $F_2(q_k,P_k)$ , with $M_1$ and $M_2$ given by (\ref{ctmat}), is much more sophisficated than (\ref{nmat}). However, for the {\it specific} $F_2$ (and  $F_3$) that we will use below, we do have 
$\partial Q_i/\partial q_j=\delta_{ij}=\partial P_i/\partial p_j$, and (\ref{ctmat}) simply reduces back to
(\ref{nmat}). 

Similarly for $F_3$: the first equation in (\ref{f3}) is an implicit
equation for determining $Q_i$ in terms of $q_i$ and $p_i$ and the
second is an explicit equation for determining $P_i$ in 
terms of $p_i$ {\it and the updated $Q_i$}. This is again sequential updating with
$\det M=1$.

Among all canonical transformations, the most important one is when
$Q_i\!=\!q_i(t)$ and $P_i\!=\!p_i(t)$, which evolves the {\it dynamics} of
the system to time $t$. Clearly, in this case, $K(Q_i,P_i)\!=\!H(q_i,p_i)$ and both the new
and old variables obey their respective Hamilton's equations. For an arbitrary
$t$, such transformation is generally unknown. However, when $t$ is infinitesimally
small, $t\rightarrow \dt$, it is well known that the Hamiltonian is the {\it infinitesimal }  
generator of time evolution.\cite{gol80, lan60} What was not realized for a long time is that
even when $\dt$ is {\it not} infinitesimally small, the resulting canonical transformation 
generated by the Hamiltonian,  through no longer exact in evolving the system in time, 
remained an excellent numerical algorithm. Let's take 
\be
F_2(q_i,P_i)=\sum_{i=1}^n q_iP_i+\dt H(q_i,P_i),
\la{f2h}
\ee
where $H(q_i,p_i)$ is given by (\ref{ham}). 
For this generating function, $\dt$ is simply an arbitrary parameter, need
not be small. 
The transformation equations (\ref{f2}) then
give,
\be
P_i=p_i-\dt\frac{\partial V(q_i)}{\partial q_i},\qquad Q_i=q_i+\dt\frac{P_i}m .
\la{alf2} 
\ee
If one regards $(Q_i,P_i)$ as $(q_i(\dt),p_i(\dt))$,
then the above is precisely Cromer's algorithm.
The transformation (\ref{alf2}) is canonical regardless of the size of $\dt$, but it is
an accurate scheme for evolving the system forward in time only when $\dt$ is
sufficiently small.
Similarly, taking
\be
F_3(p_i,Q_i)=-\sum_{i=1}^n p_iQ_i+\dt H(Q_i,p_i)
\la{f3h}
\ee
gives the other canonical algorithm
\be
Q_i=q_i+\dt\frac{p_i}m,\qquad P_i=p_i-\dt\frac{\partial V(Q_i)}{\partial Q_i},
\la{alf3}  
\ee
which is Stanley's second {\it Last Point Approximation} algorithm. \cite{sta84} 
These two are the {\it canonical} (literally and technically), first-order algorithms for evolving
classical dynamics. For our later purpose of uniformly labeling higher order methods, 
we will refer to (\ref{alf2})
and (\ref{alf3}) as algorithm 1A and 1B respectively.

When these two algorithms are iterated, they are structurally
identical to the leap-frog algorithm, which is sequential. The only difference is that,
in leap-frog, each updating of $p_i$ or $q_i$ is viewed as occurring at successive time, one
after the other. Here, $p_i$ {\it and} $q_i$ are also updated sequentially but the updating of both 
is regarded as happening at the same time.

\begin{figure}[t]
\includegraphics[width=0.49\linewidth]{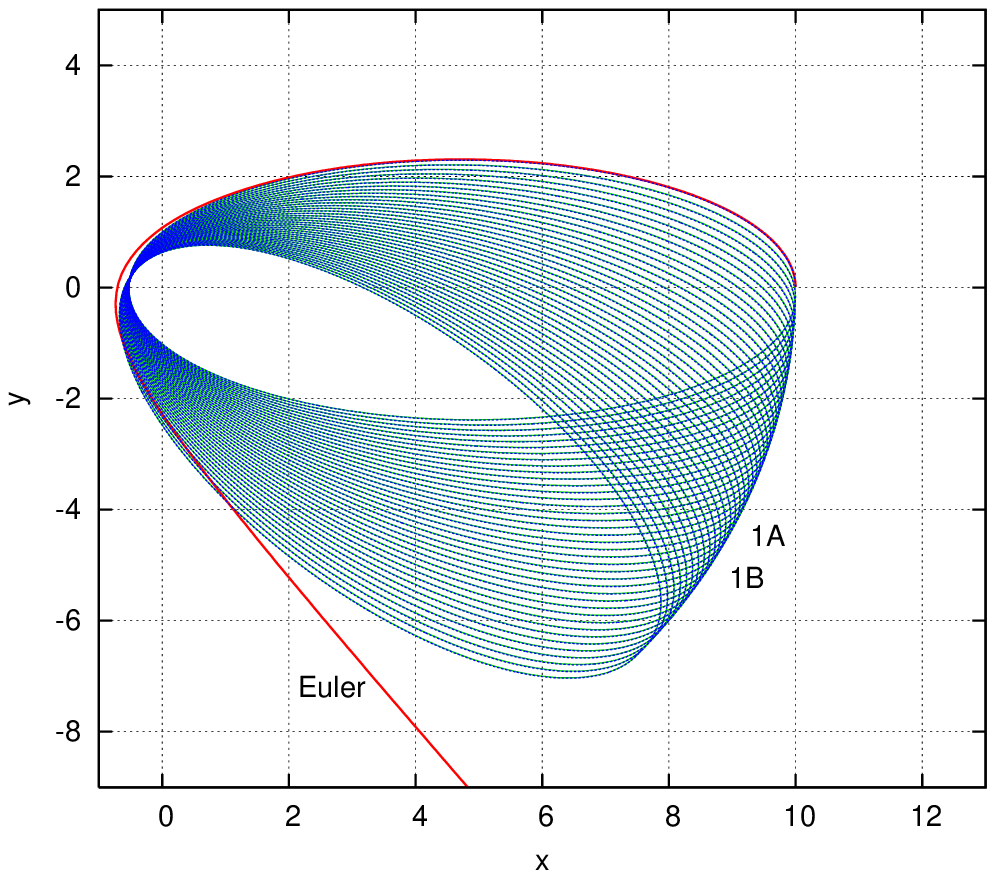} 
\includegraphics[width=0.49\linewidth]{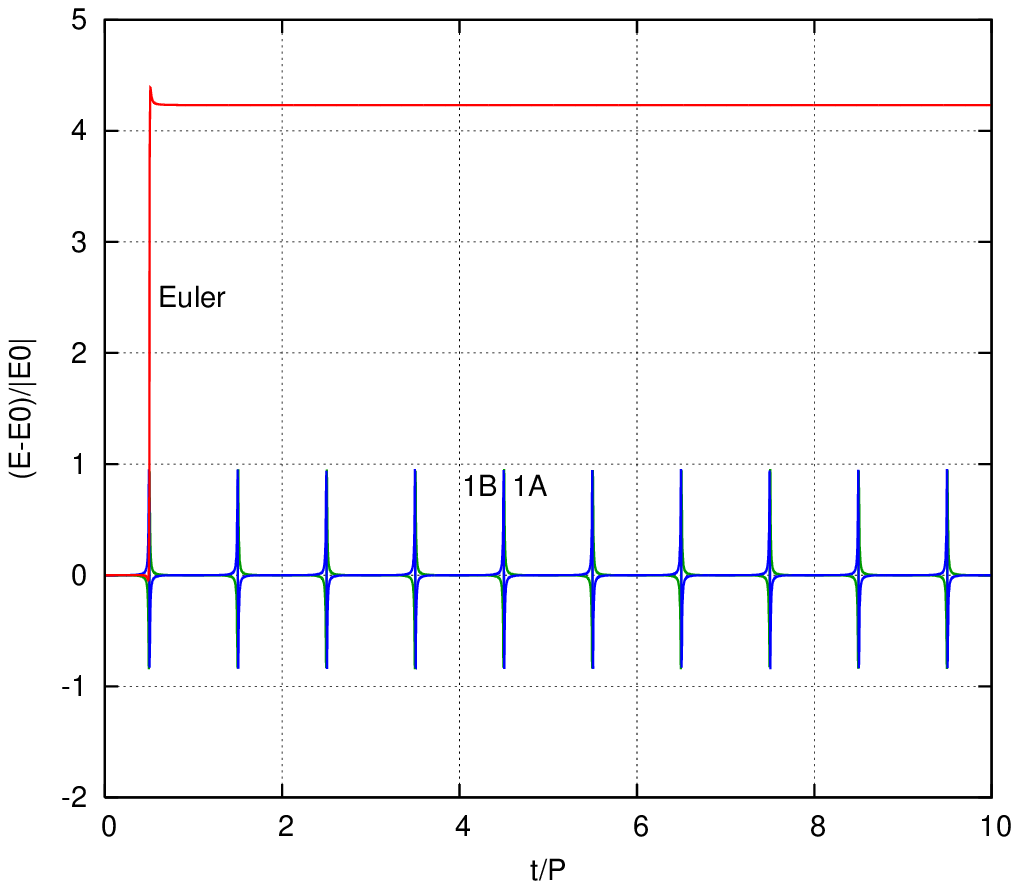} 
\caption{(Color online) First order algorithms 1A, 1B and Euler at $\dt=0.1$ {\bf Left}: orbital trajectory;
{\bf Right}: normalized energy error as a function of $t$ in unit of the period $P$ of
the exact orbit. See text for details.}
\label{eng1}
\end{figure}

\begin{figure}[t]
\includegraphics[width=0.90\linewidth]{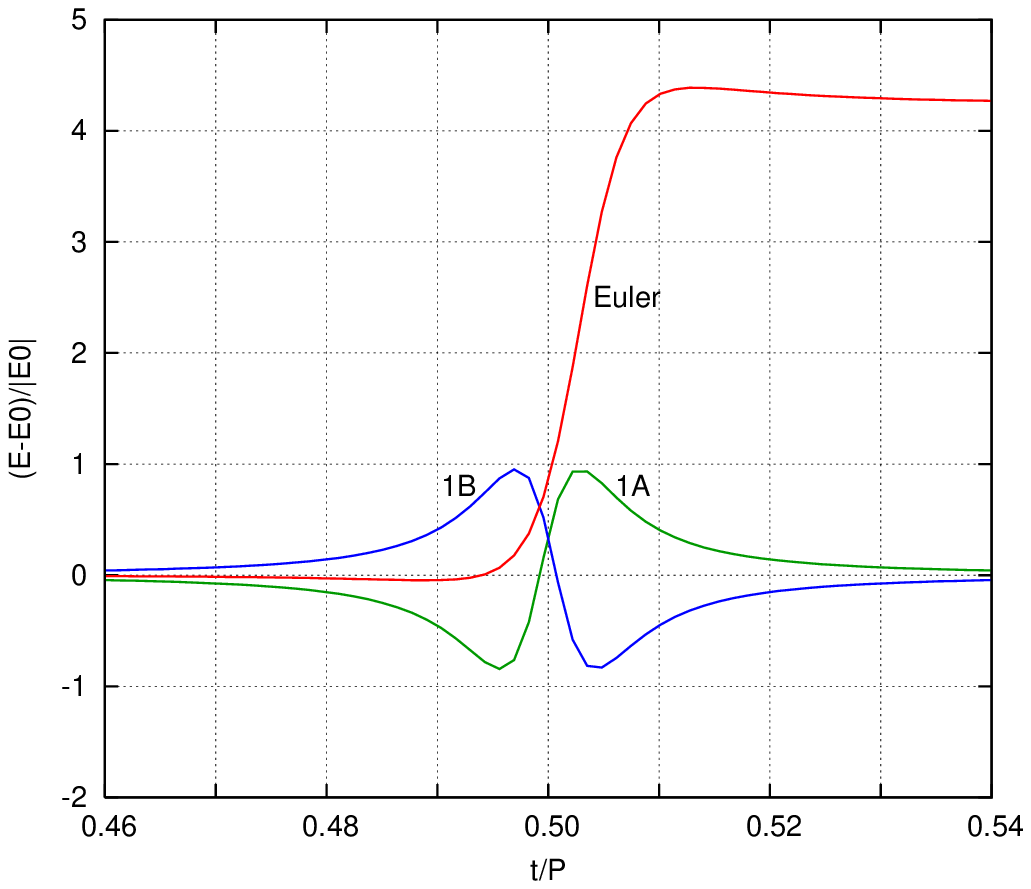} 
\caption{(Color online) Fractional energy errors of three first-order algorithms at midperiod.}
\label{eng1p1}
\end{figure}

To see how these algorithms actually perform, we solve the dimensionless
two dimensional Kepler problem defined by
$$
H=\frac12 \bp^2 -\frac{1}{|\br|} , 
$$
with initial condition $\br=(10,0)$ and $\bp=(0,0.1)$ at $\dt=0.1$.
In Fig.\ref{eng1}, we show the resulting trajectories and energy errors for
$\approx 40$ periods. Because $\dt$ has been deliberately chosen to be large, the orbits
generated by 1A and 1B do not close; they just precess. This precession of the Keplerian
orbit is a common feature of CT, corresponding to perturbing the original Hamiltonian
by the algorithm's error terms defined in the next Section. 
(This precession is similar to Mercury's perihelion advance 
due to the perturbation of general relativity.)  Since 1A and 1B 
are very similar when iterated, their orbits, green and blue respectively, overlapped to yield a
turquoise trajectory. By contrast, Euler can only complete half 
a orbit before the trajectory is ejected to infinity. On the right of Fig.\ref{eng1} are the fractional
energy errors of each algorithm. For 1A and 1B, the energy errors only spike (but at nearly 100\% !) 
{\it periodically} at midperiod. In the next Section, we will explain why the trajectory remained
fairly intact despite such huge energy errors.
In Fig.\ref{eng1p1}, we zoom in at midperiod to resolve the fine
structures. It shows clearly that 1A and 1B have equal and opposite energy errors. 

\section {Poissonian mechanics and the algebraization of algorithms}
\la{lie} 

While it is easy to use (\ref{f2h}) and (\ref{f3h}) to produce first-order algorithms, it is not
obvious how one can generalize either to obtain higher-order, more accurate methods.
For this, we turn to the Poissonian formulation mechanics in terms of Poisson brackets.

For any dynamical variable $W(q_i,p_i)$,  its evolution
through the evolution of $q_i(t)$ and $p_i(t)$ is just
\begin{eqnarray}
\frac{d}{dt} W(q_i,p_i)
          =&&\sum_{i=1}^n\Bigl({{\partial W}\over{\partial q_i}}
                  {{\partial q_i}\over{\partial t}}
				 +{{\partial W}\over{\partial p_i}}
                  {{\partial p_i}\over{\partial t}}\Bigr),\nonumber\\
          =&&\sum_{i=1}^n\Bigl({{\partial W}\over{\partial q_i}}
                  {{\partial H}\over{\partial p_i}}
				 -{{\partial W}\over{\partial p_i}}
                  {{\partial H}\over{\partial q_i}}\Bigr)\equiv\{W,H\},
\label{peq}
\end{eqnarray}
where Hamilton's equations (\ref{hameq}) have been used to obtain the
second line. The last equality defines the Poisson bracket of $W$ and $H$.
We shall refer to this as Poisson's equation-of-motion. This equation
now gives us an alternatively way of derving canonical transformations.
Let us rewrite (\ref{peq}) as
\be
{{dW }\over{dt}}=\{W,H\}=\hat H W,
\label{peqop}
\ee 
and {\it regard the Poisson bracket  $\{W,H\}$ as defining an operator $\hat H$ acting on $W$}.
Clearly then,
$$
\hat H=
\sum_{i=1}^n\Bigl(
                  {{\partial H}\over{\partial p_i}}{{\partial }\over{\partial q_i}}
				 - {{\partial H}\over{\partial q_i}}{{\partial }\over{\partial p_i}}\Bigr).
$$
For the standard Hamiltonian (\ref{ham}), this operator is just
$$
\hat H
=\sum_{i=1}^n\Bigl(
   {{p_i}\over{m}}{{\partial }\over{\partial q_i}}
              +F_i{{\partial }\over{\partial p_i}}
                  \Bigr), 
$$
which we can write it compactly as
\ba
\hat H &=& \hat T +\hat V \nn\\
&=&\bv\cdot \frac{\partial }{\partial \bq}+\ac(\bq)\cdot \frac{\partial }{\partial \bv}, 
\la{cform}
\ea
with $\bv\equiv{\bf p}/m$ and $\ac\equiv-(\partial V/\partial \bq)/m$.  
We'll just refer to $\hat H$, $\hat T$, and $\hat V$, as the classical Hamiltonian, kinetic energy and
potential energy {\it operators} respectively. 

For $H$ not explicitly dependent on time, (\ref{peqop})
can be integrated to obtain the formal operator solution
\be
W(q_i,p_i,t)
={\rm e}^{t\hat H}W(q_i,p_i,0).
\label{peqform}
\ee
Let's pause here to appreciate this remarkable Poissonian achievement. In the
Newtonian, Lagrangian or Hamiltonian formulation, only the equations-of-motion are given,
and one must solve them still. Here, in the Poissonian formulation,
the {\it solution} (\ref{peqform}) is already given, there is nothing more to solve! To see this, let's
consider the harmonic oscillator case with $\ac(\bq)=-\om^2\bq$ and apply (\ref{peqform}) to 
$\bq(t)$ and $\bv(t)$:

\ba
\left(\begin{array}{c}
       \bq(t)\\
       \bv(t)
       \end{array}\right)&=&{\rm e}^{t \hH}\left(\begin{array}{c}
       \bq\\
       \bv
       \end{array}\right)\nn\\
&=&(1+t\tH+\frac12 t^2\tH^2+\frac1{3!}t^3\tH^3+\frac1{4!}t^4\tH^4+\cdots)
\left(\begin{array}{c}
       \bq\\
       \bv
       \end{array}\right)\nn
\ea       
From (\ref{cform}), we see that
$$
\tH\bq=\bv,\quad\tH^2\bq=-\om^2\bq,\quad\tH^3\bq=-\om^2\bv,\quad\tH^4\bq=\om^4\bq\quad \cdots
$$
and therefore
\ba
\left(\begin{array}{c}
       \bq(t)\\
       \bv(t)
       \end{array}\right)&=&\left(\begin{array}{c}
       \bq+t\bv-\frac12 t^2\om^2\bq-\frac1{3!}t^3\om^2\bv+\frac1{4!}t^4\om^4\bq+\cdots\\
      \ \ \bv-t\om^2\bq-\frac12 t^2\om^2\bv+\frac1{3!}t^3\om^4\bq+\frac1{4!}t^4\om^4\bv+\cdots
       \end{array}\right)\nn\\
&=&
\left(\begin{array}{c}
       \ \ \ \cos(\om t)\bq+\om^{-1}{\sin(\om t)}\bv\\
         -{\om}\,{\sin(\om t)}\bq+\cos(\om t)\bv
       \end{array}\right)\nn
\ea       
which are the exact solutions with $\bq=\bq(0)$ and $\bv=\bv(0)$. Of course, when
$\ac(\bq)$ is not linear in $\bq$, it is generally not possible to sum the infinite series.
However, we do have exact, closed forms for
\ba
{\rm e}^{\dt \hT}\left(\begin{array}{c}
       \bq\\
       \bv
       \end{array}\right)&=&(1+\dt\bv\cdot\partial_\bq
+\frac12 \dt^2(\bv\cdot\partial_\bq)^2+\cdots)\left(\begin{array}{c}
       \bq\\
       \bv
       \end{array}\right)\nn\\
&=&\left(\begin{array}{c}
       \bq+\dt\bv\\
       \bv
       \end{array}\right),
\nn       
\ea 
and 
\ba
{\rm e}^{\dt \hV}
\left(\begin{array}{c}
       \bq\\
       \bv
\end{array}\right)
&=&\left(\begin{array}{c}
       \bq\\
       \bv+\dt\ac(\bq)
       \end{array}\right).
\nn       
\ea 
Therefore, this suggests that one should regard the classical evolution operator
${\rm e}^{t \hH}$ in (\ref{peqform}) as $[{\rm e}^{\dt (\hT+\hV)}]^n$,
and approximate the short-time evolution operator ${\rm e}^{\dt (\hT+\hV)}$
in terms of ${\rm e}^{\dt\hT}$ and ${\rm e}^{\dt\hV}$. Recalling the
well-known Baker-Campell-Hausdorff formula,\cite{mil72}
$$
{\rm e}^{A}{\rm e}^{B}=
{\rm exp}\left[A+B+{1\over 2}[A,B]+{1\over{12}}[A,[A,B]]+{1\over{12}}[B,[B,A]]+\cdots\right]
$$
where $[A,B]\equiv AB-BA$, one immediately sees that 
\be
{\cal T}_{1A}(\dt)\equiv{\rm e}^{\dt\hat V}{\rm e}^{\dt \hat T}={\rm e}^{\dt \hH_{1A}}
\la{sym1a}
\ee
with
\begin{equation}
\hH_{1A}=\hH+{1\over 2}\dt[\hV,\hT]
+{1\over{12}}\dt^2([\hV,[\hV,\hT]]+[\hT,[\hT,\hV]])\cdots.
\label{h1a}
\end{equation}
Therefore when the right most term in (\ref{sym1a}) is expanded, its first order $\dt$ term 
(for the trajectory) is correct. 
At the same time, (\ref{h1a}) shows that $\hH_{1A}$, which is the {\it exact} Hamiltonian operator
for ${\rm e}^{\dt\hat V}{\rm e}^{\dt \hat T}$, has a first-order $\dt$ error term as compared to
the original Hamiltonian operator $\hH$.  
The effect of ${\cal T}_{1A}$ is then
\ba
\e^{\dt \hV}\e^{\dt \hT}
\left(\begin{array}{c}
       \bq\\
       \bv
\end{array}\right)&=&\e^{\dt \hV}
\left(\begin{array}{c}
       \bq+\dt\bv\\\
       \bv
\end{array}\right)\nn\\
&=&
\left(\begin{array}{c}
       \bq+\dt(\bv+\dt\ac(\bq))\\
       \bv+\dt\ac(\bq)
\end{array}\right).
\nn
\ea
Again, with $\bq=\bq_0$, $\bv=\bv_0$, and the last numbered values as the updated values, 
the above corresponds to the updating
\ba
\bv_1&=&\bv_0+\dt\ac_0\nn\\
\bq_1&=&\bq_0+\dt\bv_1,
\la{al1a}
\ea
which is the previously derived algorithm 1A.
Note that the operators are applied from {\it right to left} as indicated, but the
last applied operator gives the first updating of the variable, and 
the dynamical variables are updated in the reversed order according to
the operators from {\it left to right}.

Similarily, the approximation
\be
{\cal T}_{1B}(\dt)\equiv{\rm e}^{\dt\hat T}{\rm e}^{\dt \hat V}={\rm e}^{\dt \hH_{1B}}
\la{sym1b}
\ee
with
\begin{equation}
\hH_{1B}=\hH+{1\over 2}\dt[\hT,\hV]
+{1\over{12}}\dt^2([\hT,[\hT,\hV]]+[\hV,[\hV,\hT]])\cdots.
\label{h1b}
\end{equation}
gives
\ba
\e^{\dt \hT}\e^{\dt \hV}
\left(\begin{array}{c}
       \bq\\
       \bv
\end{array}\right)=
\e^{\dt \hT}
\left(\begin{array}{c}
       \bq\\
       \bv+\dt\ac(\bq)
\end{array}\right)
&=&
\left(\begin{array}{c}
       \bq+\dt\bv\\
       \bv+\dt\ac(\bq+\dt\bv)
\end{array}\right).
\nn
\ea
corresponds to the updating
\ba
\bq_1=\bq_0+\dt\bv_0\nn\\
\bv_1=\bv_0+\dt\ac_1,
\la{al1b}
\ea
where $\ac_1\equiv \ac(\bq_1)$, which is algorithm 1B. 

This way of deriving these two algorithms makes it very clear that
the algorithms are sequential because the operators are applied sequentially,
with each operator only acting on one dynamical variable at a time. More importantly,
from (\ref{sym1a}) and (\ref{h1a}), the trajectory produced by 1A is {\it exact} for the {\it modified} Hamiltonian operator $\hH_{1A}$, which approaches the original $\hH$ only as $\dt\rightarrow 0$. To find 
the modified Hamiltonian function $H_{1A}$ corresponding to $\hH_{1A}$, we recall that
$\hH W=\{W,H\}$, and therefore
\ba
[\hT,\hV]\,\,W
&=&(\hT\hV-\hV\hT)W=\hT\{W,V\}-\hV\{W,T\}\, ,\nn\\
&=&\{\{W,V\},T\}-\{\{W,T\},V\}\, 
=\{W,\{V,T\}\}\, ,
\la{comeq}
\ea
where the last equality follows from the
Jacobi identity
$$
\{\{W,V\},T\}+\{\{T,W\},V\}+\{\{V,T\},W\}=0\, .
$$
From (\ref{comeq}) we see that the function corresponding to $[\hT,\hV]$ is 
$\{V,T\}=-\{T,V\}$, and that for the separable Hamiltonian (\ref{ham})
$$
\{T,V\}=\sum_{i=1}^n-\frac{\pa T}{\pa p_i}\frac{\pa V}{\pa q_i}=\bv\cdot\bF,
$$
and therefore
\be
H_{1A} = H+\frac12 \dt\bv\cdot\bF+\cdots \quad{\rm and}\quad 
H_{1B} = H-\frac12 \dt\bv\cdot\bF+\cdots
\la{hab}
\ee
These modified Hamiltonians for algorithms 1A and 1B then explain the observations in
Fig.\ref{eng1p1}: 1) The energy errors are opposite in sign. 2) Since $|\bF|\approx 1/r^2$
and $|\bv|\approx 1/r$, the energy errors are nearly zero far from the force center. 
3) Since at the pericenter $\bv\cdot\bF=0$, the
first order energy errors cross zero at midperiod. The fact that they don't in Fig.\ref{eng1p1}
simply means that the remaining higher order error terms in (\ref{hab}) do not vanish at the pericenter. 
 For example,
$$
[\hV,[\hT,\hV]]\lra\{[\hT,\hV],V\}\lra\{\{V,T\},V\}\lra\{V,\{T,V\}\},
$$
and
\be
\{V,\{T,V\}\}=\sum_{i=1}^n\frac{\pa V}{\pa q_i}\frac{\pa \{T,V\}}{\pa p_i}
=-\frac{\bF\cdot\bF}{m},
\la{potvtv}
\ee
which peaks strongly near the force center as $1/|\br|^4$. 4) The error terms in (\ref{hab}), spoil the
the exact inverse-square nature of the force, and therefore by Bertrand's theorem,\cite{ber73,chin15} the orbit cannot close, resulting in precessions as seen in Fig.\ref{eng1}. 5) Because of this precession, the energy
errors are periodic at midperiod.
We can explain these observations in details because a symplectic integrator remains governed by a Hamiltonian, though modified. Very little can be say about the
behavor of non-symplectic algorithms such as Euler, except that they are ultimately unstable because
they are not bound by a Hamiltonian.

In general then, one can seek to approximate ${\rm e}^{\dt (\hT+\hV)}$ by a
product of ${\rm e}^{\dt \hT}$ and ${\rm e}^{\dt \hV}$ in the form of 
\be
{\rm e}^{\dt (\hT+\hV)}=\prod_{i=1}^{n}{\rm e}^{c_i\dt \hT}{\rm e}^{d_i\dt \hV},
\la{prod}
\ee
with real coefficients $\{c_i,d_i\}$ determined to match ${\rm e}^{\dt (\hT+\hV)}$
to any order in $\dt$. Since a product of operators represent a sequential updating
of one dynamical variable at a time with Jacobians (\ref{nmat}), 
the resulting algorithm is guranteed to have $\det M=1$. 
Also, since each operator is a canonical transformation, the entire sequence is also a canonical transformation.
Thus we now have a method of generating arbitrarily accurate canonical transformations.
Moreover, this process of {\it decomposition}, of approximating ${\rm e}^{\dt (\hT+\hV)}$ by a
product of ${\rm e}^{c_i\dt \hT}$ and ${\rm e}^{d_i \dt \hV}$, 
reduces the original problem of Taylor series {\it analysis} into an {\it algebraic}
problem of operator factorization. Finally, this product
approximation of ${\rm e}^{\dt (\hT+\hV)}$, for any two operators $\hT$ and $\hV$, can be applied to  
any other physical evolution equation. 

Similar to the Hamiltonian function, one can define an operator $\hat S$ corresponding to any dynamical function $S(q_i,p_i)$ via the Poisson bracket $\{W,S\}=\hat S W$. 
It is then easy to prove\cite{dep69,boc98} that the transformation generated by
$\e^{\epsilon \hat S}$, for any $S(q_i,p_i)$, is also canonical. Such an operator $\hat S$ is called
a {\it Lie operator} and the series evaluation of $\e^{\epsilon \hat S}$ as 
the {\it Lie series}\cite{dep69} method. These seminal ideas from
celestial mechanics\cite{dep69,boc98} and accelerator physics, \cite{dra76,fr90,for06} forged
the modern development of symplectic integrators.\cite{ner87,yos93}
(In this Section, it is also imperative that we use an intuitive and non-intimidating notation
to denote Lie operators such as $\hat H$ or $\hat S$.
The direct use\cite{gol80} of $\{\cdot,S\}$ leads to unwieldy expression for powers of $\hat S^n$ 
as $...\{\{\{\{\cdot,S\},S\},S\},S\}..$. The old time typographical notation\cite{dra76,fr90} of
``:$S$:'' is unfamiliar to modern readers. The use of ``$L_S$'' \cite{dep69,boc98} or ``$D_S$'' \cite{don05} is redundant, with the symbol ``$L$'' or ``$D$'' serving no purpose. 
In this work, we eschew these three conventions and 
follow a practice already familiar to students (from quantum mechanics) by denoting
a Lie operator with a ``caret'' over its defining function, $\hat S$. Also,
it is essential to define $\hat H=\{\cdot,H\}$, and {\it not} $\hat H=\{H,\cdot \}$,
otherwise one gets a confusing negative sign for moving forward in time.)

\section {Time-reversibility and second-order algorithms}
\la{trh}

The two first-order algorithms derived so far, despite being cannonical,
are {\it not} time-reversible, that is, unlike the exact trajectory,
they cannot retrace their steps back to the original starting point when run backward in time. This point
is not obvious when they were first derived from generating functions. In the operator formulation,
this is now transparent because
$$
\cT_{1A}(-\dt)\cT_{1A}(\dt)=\e^{-\dt \hV}\e^{-\dt \hT}\e^{\dt \hV}\e^{\dt \hT}\neq 1.
$$
Similarly for $1B$. This means that these two first-order algorithms may yield trajectories
qualitatively {\it different}  from the exact solution. For example, the phase trajectory for the harmonic 
oscillator with $H=(p^2+q^2)/2$ ($\om=1$) should just be a circle. However, the modified Hamiltonians for 1A and 1B from (\ref{hab}) are 
$$
H_{1A} = \frac12(p^2+q^2-\dt p q)
=\frac12\left( \frac{u^2}{(1-\dt/2)^{-1}}+ \frac{v^2}{(1+\dt/2)^{-1}} \right),
$$
$$
H_{1B} =\frac12(p^2+q^2+\dt p q)
=\frac12\left( \frac{u^2}{(1+\dt/2)^{-1}}+ \frac{v^2}{(1-\dt/2)^{-1}} \right),
$$
with $u=(p+q)/\sqrt{2}$ and $v=(p-q)/\sqrt{2}$. For $|\dt|<2$, these are now 
ellipses\cite{lar83,don05} with semi-major axes tilted at $\pm 45^\circ$. This $45^\circ$ tilt is
uncharacteristic of the exact trajectory.

This shortcoming can be easily fixed by just cancatenating $1A$ and $1B$ together, giving
\be
{\cal T}_{2A}(\dt)=\cT_{1A}(\dt/2)\cT_{1B}(\dt/2)
=\e^{\dt \hV/2}\e^{\dt \hT}\e^{\dt \hV/2}=\e^{\dt \hH_{2A}(\dt)},
\la{sym2a}
\ee
with left-right symmetric operators, guaranteeing time-reversibility:
\be
{\cal T}_{2A}(-\dt){\cal T}_{2A}(\dt)=\e^{-\dt \hV/2}\e^{-\dt \hT}\e^{-\dt \hV/2}\e^{\dt \hV/2}\e^{\dt \hT}\e^{\dt \hV/2}=1.
\la{tr}
\ee
This means that the algorithm's Hamiltonian operator must be of the form
\be
\dt\hH_{2A}(\dt)=\dt(\hH+\dt^2 \hat E_2+\dt^4 \hat E_4+ \cdots) ,
\la{dth2a}
\ee
with only even-order error operators $\hat E_{2},\hat E_{4},$ etc., otherwise, 
if $\hH_{2A}(\dt)$ has any odd-order error term, then $\dt \hH_{2A}(\dt)$
would have a term even in $\dt$ and
$
\e^{-\dt \hH_{2A}(-\dt)}\e^{\dt \hH_{2A}(\dt)}\neq 1,
$
contradicting (\ref{tr}). {\it Thus any decomposition with left-right symmetric operators
must yield an even-order algorithm, of at least second-order}. 

\begin{figure}[t]
\includegraphics[width=0.49\linewidth]{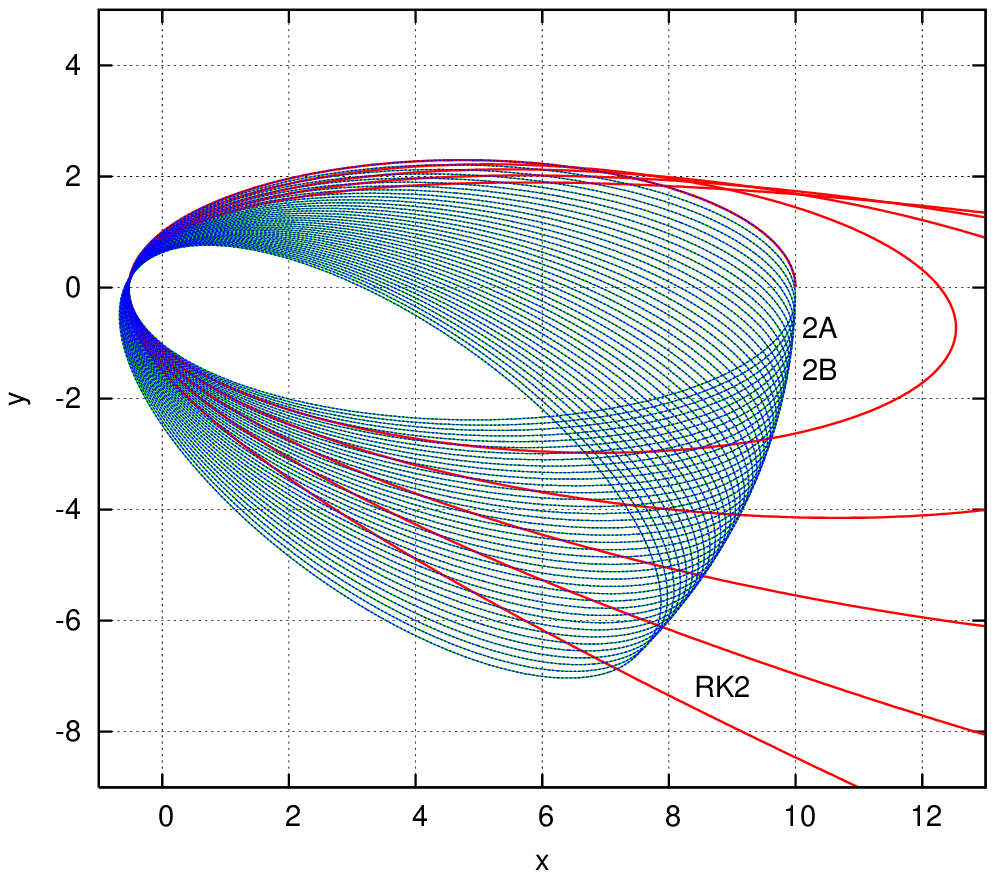} 
\includegraphics[width=0.49\linewidth]{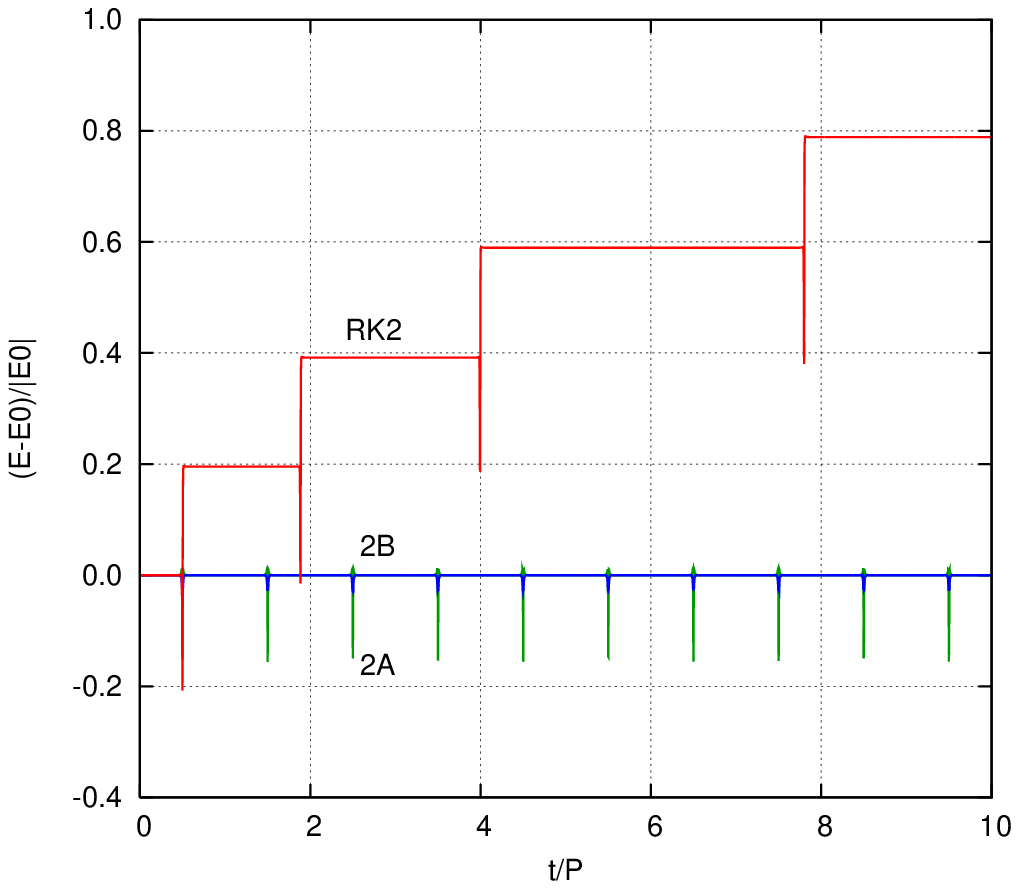} 
\caption{(Color online) Trajectory and energy errors similar to Fig.\ref{eng1} for three second-order algorithms 2A, 2B and RK2 at $\dt=0.1$ .}
\label{eng2}
\end{figure}

\begin{figure}[t]
\includegraphics[width=0.95\linewidth]{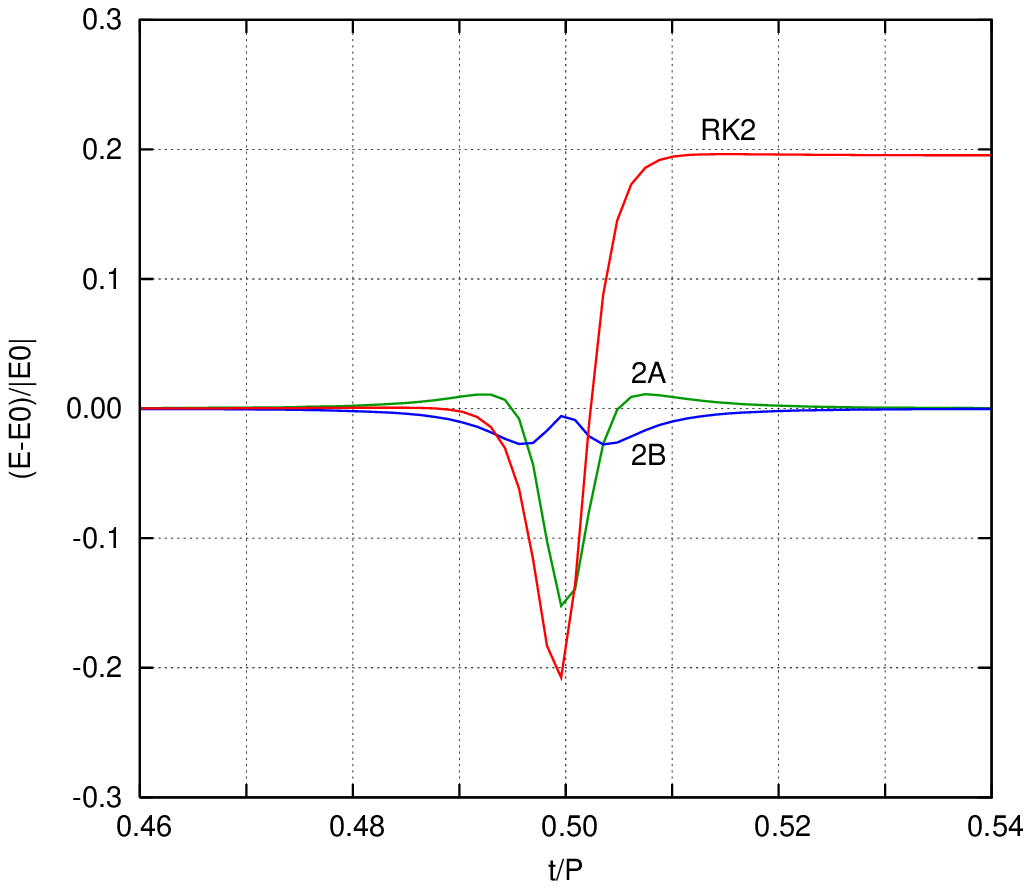} 
\caption{(Color online) Energy errors at midperiod for the three second-order 
algorithms of Fig.\ref{eng2}.}
\label{eng1p2}
\end{figure}

The action of ${\cal T}_{2A}(\dt)$ 
\ba
\e^{\dt \hV/2}\e^{\dt \hT}\e^{\dt \hV/2}
\left(\begin{array}{c}
       \bq\\
       \bv
\end{array}\right)&=&
\e^{\dt \hV/2}
\left(\begin{array}{c}
       \bq+\dt\bv\\
       \bv+\frac12 \dt\ac(\bq+\dt\bv)
\end{array}\right)\nn\\
&=&
\left(\begin{array}{c}
       \bq+\dt\Bigl[\bv+\frac{\dt}{2} \ac(\bq)\Bigr]\\
       \Bigl[\bv+\frac{\dt}{2} \ac(\bq)\Bigr]+\frac12 \dt\ac(\bq+\dt\Bigl[\bv+\frac{\dt}{2} \ac(\bq)\Bigr]),
\end{array}\right),
\nn
\ea
yields the second-order algorithm 2A:  
\ba
\bv_1&=&\bv_0+\frac12 \dt \ac_0,\nn\\
\bq_1&=&\bq_0+\dt \bv_1,\nn\\
\bv_2&=&\bv_1+\frac12 \dt \ac_1.
\la{al2a}
\ea
Similarly,
cancatenating 1B and 1A together gives 
\be
{\cal T}_{2B}(\dt)=\cT_{1B}(\dt/2)\cT_{1A}(\dt/2)
=\e^{\dt \hT/2}\e^{\dt \hV}\e^{\dt \hT/2}=\e^{\dt \hH_{2B}(\dt)},
\la{sym2b}
\ee 
yielding the corresponding 2B algorithm
\ba
\bq_1&=&\bq_0+\frac12 \dt \bv_0,\nn\\
\bv_1&=&\bv_0+\dt \ac_1,\nn\\
\bq_2&=&\bq_1+\frac12 \dt \bv_1.
\la{al2b}
\ea
These two symplectic schemes, when solving the harmonic oscillator, do not have any cross term 
$\propto p q$ in their modified Hamiltonians, and therefore will not have any extraneous $45^\circ$ tilts in their phase trajectories.

The sequential steps of algorithms 2A and 2B can be directly translated into programming lines 
and there is no need to combine them. However, for comparison, steps in 2A can be combined to give 
\ba
\bq_1&=&\bq_0+\dt \bv_0+\frac12 \dt^2 \ac_0,\la{qone}\\
\bv_2&=&\bv_0+\frac12 \dt (\ac_0+ \ac_1),
\la{vv}
\ea
which is just the velocity-Verlet\cite{gou07} algorithm, or Stanley's {\it second-order Taylor
approximation}, STA.\cite{sta84} Note that 
the acceleration computed at the end, $\ac_1$, can be reused as $\ac_0$ 
at the start of the next iteration. Thus for both algorithms, only one force-evaluation 
is needed per iteration.
Algorithm 2B is referred to as the position-Verlet\cite{gou07} algorithm,
which corresponds to Stanley's STA$^\prime$ algorithm. 
These two {\it canonical} second-order algorithms were known before Stanley's
time, but were unrecognized as symplectic integrators among a host of other 
similarly looking algorithms.\cite{sta84}

We compare the working of these two second-order algorithm in Fig.\ref{eng2}. Not suprisingly,
the lengthy iteration of 2A and 2B yielded trajectories similar to those of 1A and 2B. Note however,
that their periodic energy errors are now much smaller. As shown in Fig.\ref{eng1p2}, the 
maximum energy errors of 2A and 2B are only $\approx 15\%$ and $\approx 3\%$ respectively.
Since the expected errors are on the order of O($\dt^2$), 2B's error is entirely reasonable,
while that of 2A is larger than expected. Nevertheless, these errors are more commensurate with
their well-behave trajectories, similar to those of 1A and 1B. Thus
the trajectories of 1A and 1B are nearly second-order; 
only that their energy errors do not reflect this.
Plotted also are results for a Runge-Kutta type second-order algorithm RK2 derived in
Section \ref{rkn}. Its trajectory manages to return to the force center a few times, but was eventually
ejected to infinity. As we will see, it is characteristic of Runge-Kutta algorithms that the magnitude of their energy errors is {\it non-periodic} and ever increasing with each orbit.

\section {Higher order symplectic algorithms}
\la{hisym}

We have derived two second-order
algorithms effortlessly without doing any Taylor expansions. 
We will now show, with
equal ease, how to derive higher order symplectic algorithms.
From (\ref{dth2a}), one has (where $\ct_2$ denotes
either $\ct_{2A}$ or $\ct_{2B}$)
\begin{equation}
{\cal T}_{2}(\dt)=\exp(\dt \hat H+\dt^3 \hat E_2+\dt^5 \hat E_4+\cdots),
\label{tb}
\end{equation}
with only odd powers of $\dt$ in the exponent. If one were able to get rid of the
$\dt^3$ term, one would have a fourth-order algorithm. 
Consider then a triple product of ${\cal T}_{2}$ with a negative time step $-s\ep$ at the middle:
\begin{equation}
{\cal T}_{2}(\epsilon)
{\cal T}_{2}(-s\epsilon)
{\cal T}_{2}(\epsilon)
={\rm exp}[(2-s)\epsilon \hat H+(2-s^3)\epsilon^3\hat E_2
+O(\epsilon^5)].
\label{tripd}
\end{equation}
This algorithm evolves the system forward 
for time $\epsilon$, {\it backward} for time
$s\epsilon$ and forward
again for time $\epsilon$. Since it is manifestly left-right symmetric,
and hence time-reversible, its exponent must consist of terms with only 
odd powers of $\epsilon$. Moreover, its leading $\ep^3$ term can only be due
to the sum of $\ep^3$ terms from each constituent algorithm, without
any commutator terms. This is because any single-commutator of $\ep\hat H$, 
$\ep^3\hat E_2$, etc., would be even in $\ep$ and hence excluded. Any double-commutator 
would have a minimum of two $\ep\hat H$ and one $\ep^3\hat E_2$, which is already 
fifth order in $\ep$.

To obtain a fourth-order algorithm, it is only necessary to eliminate the $\ep^3$ error
term in (\ref{tripd}) by choosing
$$
s=2\,^{1/3}
$$
and scale $\epsilon$ 
back to the standard step size by setting $\dt=(2-s)\epsilon$,
\begin{equation}
{\cal T}_{4}(\dt)\equiv
{\cal T}_{2}\Bigl({\dt\over{2-s}}\Bigr)
{\cal T}_{2}\Bigl({-s\, \dt\over{2-s}}\Bigr)
{\cal T}_{2}\Bigl({\dt\over{2-s}}\Bigr)
={\rm exp}[\,\dt\hat H
+O(\dt^5)].
\label{al4}
\end{equation}
This fourth-order algorithm only requires 3 force-evaluations.

\begin{figure}[t]
\includegraphics[width=0.49\linewidth]{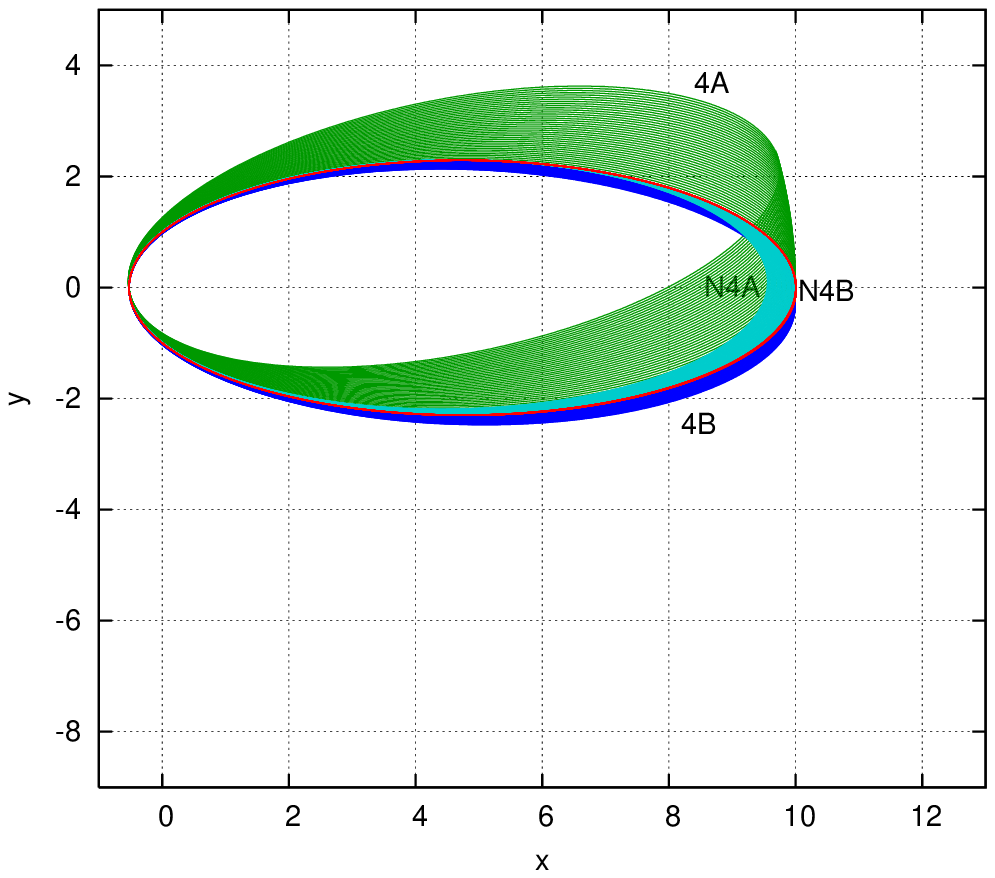}
\includegraphics[width=0.49\linewidth]{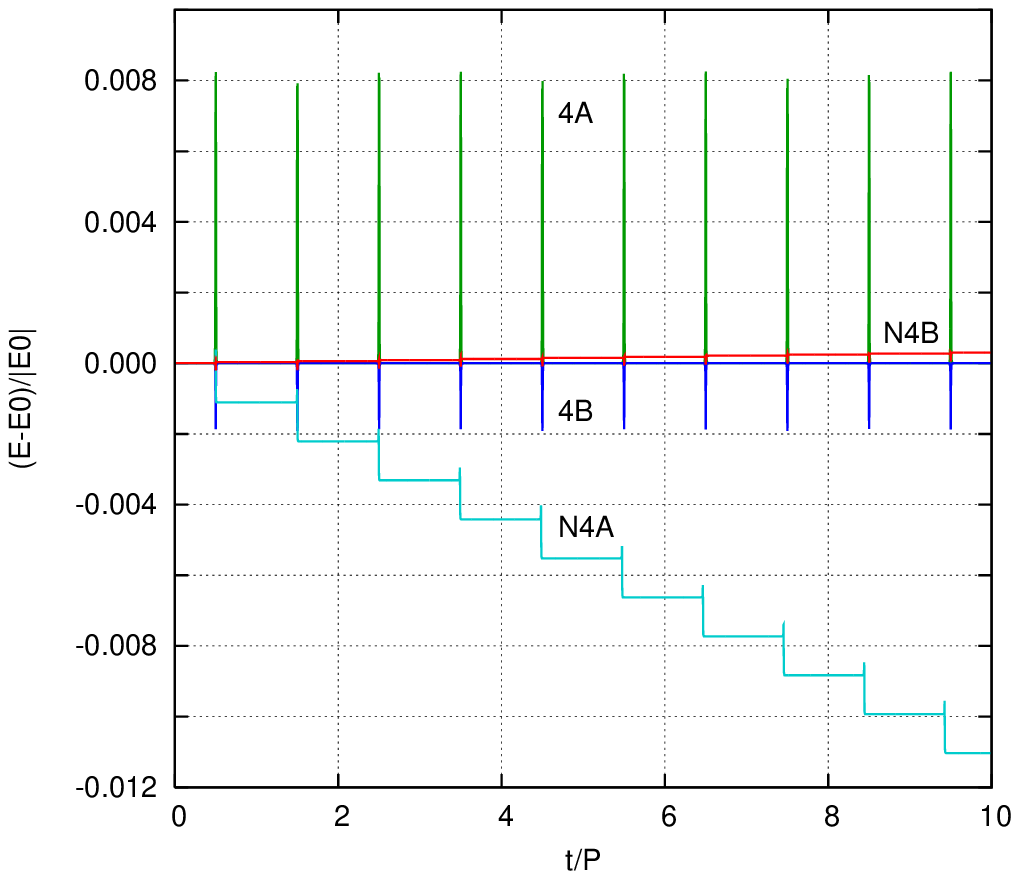} 
\caption{(Color online) Trajectory and energy errors similar to Fig.\ref{eng1} for four fourth-order algorithms 4A(green), 4B(blue), N4A(blue-green) and N4B(a red closed ellipse) at $\dt=0.1$}
\label{eng4}
\end{figure}

\begin{figure}[t]
\includegraphics[width=0.95\linewidth]{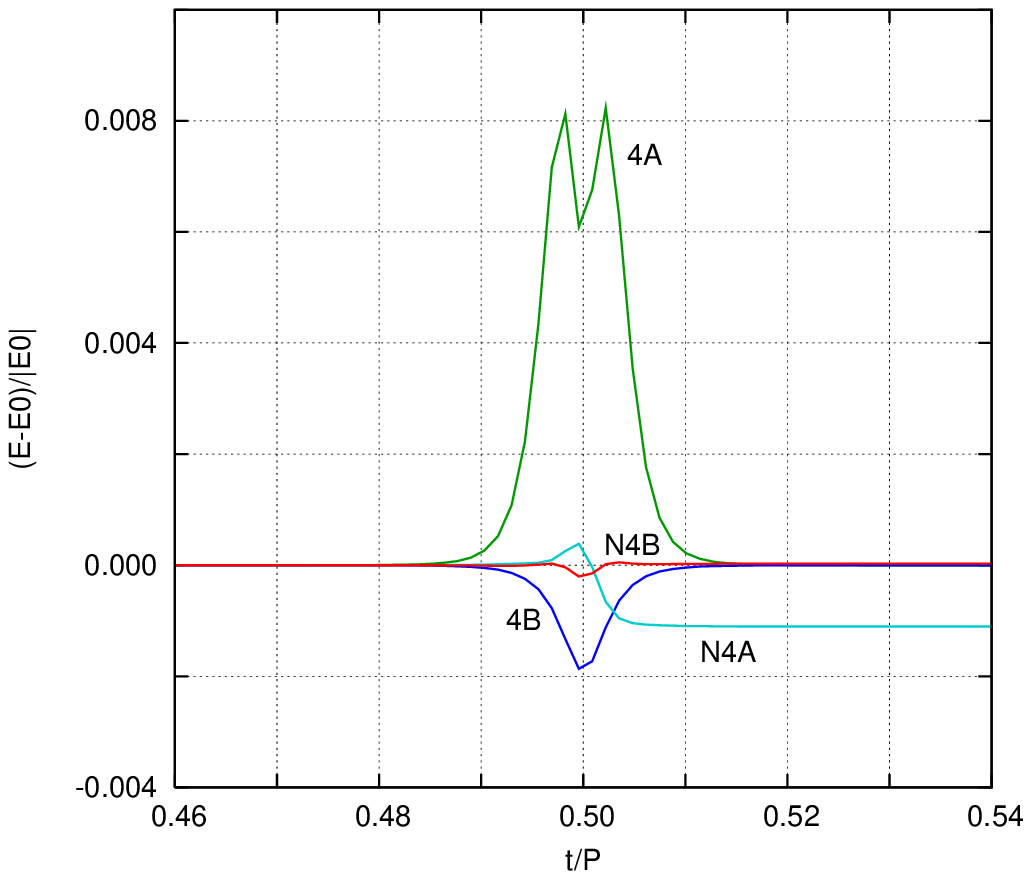}
\caption{(Color online) Energy errors at midperiod for four fourth-order 
algorithms of Fig.\ref{eng4}.}
\label{eng1p4}
\end{figure}

The derivation of the first fourth-order SI in the late 1980's
initiated the modern development of symplectic integrators. 
According to Forest,\cite{for06} the first fourth-order SI was obtained by
Ruth in 1985 by numerically solving equations with hundreds of terms.
However, their lengthy derivation of the algorithm involving a complicated
cube root was not published jointly until 1990.\cite{fr90} 
Around the same time,
many groups, including Campostrini and Rossi,\cite{cam90} Suzuki,\cite{suz90} 
Yoshida,\cite{yos90}, Candy and Rozmous\cite{can91} and Bandrauk and Shen,\cite{ban91} 
have independently arrived at the same algorithm, but it 
was Creutz and Gocksch\cite{cre89} who managed to publish it in 1989.
Here, we follow the simpler derivation of Ref.\onlinecite{chin00}.

By letting $\ct_2$ be either $\ct_{2A}$ or $\ct_{2B}$ in (\ref{al4}), one obtains
fourth-order algorithms 4A and 4B respectively. Their trajectories for the same Keplerian orbit
at $\dt=0.1$ are shown in Fig.\ref{eng4}. The orbital precession is now much reduced and their
periodic energy errors in Fig.\ref{eng1p4} are roughly ten times smaller than those of 2A and 2B.
Notice that, just as the energy error of 2B is much smaller than 2A, 4B's maximum 
energy error and orbital precession are much less than those of 4A. Also plotted are results for
Runge-Kutta type Nystr\"om algorithms N4A and N4B, which are derived in 
the next section.
   
Clearly, the triple-product construction can be repeated to generate
arbitrary even-order algorithms. For example, a sixth-order algorithm
can be obtained by taking
\begin{equation}
{\cal T}_{6}\equiv
{\cal T}_{4}\Bigl({\dt\over{2-s}}\Bigr)
{\cal T}_{4}\Bigl({-s\, \dt\over{2-s}}\Bigr)
{\cal T}_{4}\Bigl({\dt\over{2-s}}\Bigr).
\label{al6}
\end{equation}
and choosing $s=2^{1/5}$ to eliminate the fifth-order error term in ${\cal T}_{4}$.
One can then cancatenates three ${\cal T}_{6}$ to give ${\cal T}_{8}$ by
choosing $s=2^{1/7}$ etc.. Thus, without doing any tedious analysis,
one can easily generate any even-order algorithms. The disadvantage of this class of
schemes is that the the number of force-evaluation needed at order $2n$ 
grows exponentially as $3^{n-1}$. It is already uneconomical at order six, which
required nine force-evaluations. As we will see in the next section, 
a sixth-order RKN algorithm only requires five force-evaluations. 
Moreover, as shown in Fig.\ref{eng4}, even the better symplectic 4B algorithm
seems inferior, in the short run, to N4B.

\section{Higher-order Runge-Kutta-Nystr\"om algorithms}
\la{rkn}

Runge-Kutta type schemes generalize the Euler
algorithm to higher order. For example, keeping the Taylor expansion to second-order, gives
\ba
\bq(\dt)&=&\bq(0)+\dt\dot\bq(0)+\frac12 \dt^2\ddot\bq(0)\nn\\
\bq_1&=&\bq_0+\dt\bv_0+\frac12 \dt^2\ac_0
\la{rkq}
\ea
and
\ba
\bv(\dt)&=&\bv(0)+\dt\dot\bv(0)+\frac12 \dt^2\ddot\bv(0)\nn\\
\bv_1&=&\bv_0+\dt\ac_0+\frac12 \dt^2\dot\ac_0
=\bv_0+\dt(\ac_0+\frac12 \dt\bv_0\cdot\nabla\ac_0)\nn\\
&=&\bv_0+\dt\ac(\bq_0+\frac12 \dt\bv_0)+O(\dt^3)\la{rkv}\\
&=&\bv_0+\frac12 \dt[\ac_0+\ac(\bq_0+\dt\bv_0)]+O(\dt^3)
\la{rkvv}
\ea
The idea is to replace higher order time-derivatives by evaluating the force at some
intermediate positions. Both (\ref{rkv}) and (\ref{rkvv}) are correct to $\dt^2$ and are acceptable
implementation of this second-order Runge-Kutta algorithm (RK2). Both also require two evaluations of 
the force, which are inferior to only a single evaluation needed in 2A and 2B. If one chooses the
implemention (\ref{rkvv}), then one can easily check that this RK2 is just the average of   
algorithms 1A and 1B, that is
$$
{\cal T}_{RK2}(\dt)=\frac12 (
\e^{\dt \hT}\e^{\dt \hV}+\e^{\dt \hV}\e^{\dt \hT}).
$$
Thus Runge-Kutta type algorithms can also have an operator representation, except that they
are no longer a {\it single} product as in (\ref{prod}). This automatically implies that they are no longer
sequential and can't have $\det M=1$. As shown in Fig.\ref{eng2}, RK2's energy error increases
indefinitely with each period until total destablization.

The triplet construction of higher order symplectic algorithms is based on eliminating the errors
(\ref{tb}) of $\ct_{2}$, , via a product of $\ct_{2}$'s. However, since
\be
{\cal T}_{2}^k(\dt/k)=\exp(\dt \hat H+k^{-2}\dt^3 \hat E_2+k^{-4}\dt^5 \hat E_4+\cdots),
\label{tpk}
\ee
these higher order errors can alternatively be eliminated by {\it extrapolation}. For example,
a fourth-order algorithm can also be obtained by the combination
\be
{\cal T}_4(\dt)=
\frac13\Bigl[4{\cal T}_2^2\left(\frac\dt{2}\right)-{\cal T}_2(\dt)\Bigr],
\la{four}
\ee
where the factor 4 cancels the $\dt^3$ error term and the factor 3 restores
the original Hamiltonian term $\dt H$. Because this is no longer a single product
of operators, the algorithm is again no longer symplectic with $\det M=1$. Take $\ct_2$ to be $\ct_{2A}$, 
whose algorithmic form is given by (\ref{vv}), then $\ct^2_{2A}(\dt/2)$ corresponds to iterating that twice
at half the step size:
\ba
\bq_1^\prime&=&\bq_0+\frac12 \dt \bv_0+\frac18 \dt^2 \ac_0,\la{qonep}\\
\bv_2^\prime&=&\bv_0+\frac14 \dt (\ac_0+ \ac_{1^\prime}),\nn\\
\bq_2&=&\bq_1^\prime+\frac12 \dt \bv^\prime_2+\frac18 \dt^2 \ac_{1^\prime},\nn\\
\bv_3&=&\bv_2^\prime+\frac14 \dt (\ac_{1^\prime}+ \ac_2)\nn.
\ea
The first two lines are just (\ref{vv}) at $\dt/2$, and the second
two lines simply repeat that with $\bq_1^\prime$ and $\bv_2^\prime$ as initial values.
Eliminating the intermediate values $\bq_1^\prime$ and $\bv_2^\prime$ gives
\ba
\bq_2&=&\bq_0+\dt \bv_0+\frac14 \dt^2 (\ac_0+\ac_{1^\prime}),\la{qtwo}\\
\bv_3&=&\bv_0+\frac14 \dt (\ac_0+2\ac_{1^\prime}+ \ac_2).\nn
\ea
The combination (\ref{four}) then gives
\ba
\bq(\dt)&=&\frac43\bq_2-\frac13\bq_1=\bq_0+\dt\,\bv_0+\frac{\dt^2}6\left(\bac_0+2\bac_{1^\prime}\right),\la{rkn4q}\\
\bv(\dt) &=&\frac43\bv_3-\frac13\bv_2=\bv_0+\frac{\dt}6
\left(\bac_0+4\bac_{1^\prime}+2\bac_{2}-\bac_{1}\right).
\la{rkn4}
\ea
This appears to required 4 force-evaluations at $\bq_0$, $\bq_1^\prime$ (\ref{qonep}), $\bq_2$ (\ref{qtwo})
and $\bq_1$ (\ref{rkq}). However, the force subtraction above can be combined:
\ba
2\bac_{2}-\bac_{1}
&=&2\bac(\bq_2)-\bac(\bq_{1})
=2\bac(\bq_1+\delta\bq)-\bac(\bq_{1})\nn\\
&=&\bac(\bq_{1})+2\delta\bq\cdot\nabla\bac(\bq_{1})+O(\delta\bq^2)
=\bac(\bq_{1}+2\delta\bq)+O(\delta\bq^2)\nn\\
&=&\bac(\bq_0+\dt\bv_0+\frac{1}{2}\dt^2\bac_{1^\prime})+O(\dt^6),
\la{fsub}
\ea
since
\ba
\delta\bq&=&\bq_2-\bq_1=\frac{\dt^2}4 (\bac_{1^\prime}-\bac_0)\approx O(\dt^3).\nn
\ea
Hence, correct to $O(\dt^4)$, the two force-evaluations $2\bac_{2}-\bac_{1}$ in (\ref{rkn4}) can be replaced by the single force-evaluation (\ref{fsub}). The resulting algorithm, (\ref{rkn4q}) and (\ref{rkn4}), then reproduces Nystr\"om's 
original fourth-order integrator with only three force evaluations.\cite{nys25,bat99} We denote this algorithm as N4A. Using ${\cal T}_{2B}$ in (\ref{four}) yields the algorithm N4B. In Fig.\ref{eng4}, we
see that N4A's orbit is not precessing, but is continually shrinking, reflecting its continued loss of
energy after each period. N4B's energy error is monotonically increasing after each period, and will
eventually destablize after a long time. However, its increase in energy error is
remarkable small in the short term (see Fig.\ref{eng1p4}), maintaining a closed orbit 
even after 40 periods. (A clearer picture of N4B's orbit is given in Fig.\ref{eng4f} below.) 
Thus in the short term, algorithms 4A and 4B, despite being symplectic, 
are inferior to N4B. We will take up this issue in the next section.

The standard fourth-order Runge-Kutta (RK4) algorithm,\cite{wil92} capable of solving the a general position-, velocity- and time-dependent force with $\bac(\bq,\bv,t)$, requires four force-evaluations. When the force does not depend on \bv, Nystr\"om type algorithms only need three force-evaulations. When applied to Kepler's problem, RK4, despite having four force-evaluations, behaved very much like N4A, 
and was therefore not shown separately.

The extrapolation, or the multi-product expansion (\ref{four}), can be generalized to arbitrary even\cite{chin10} or odd\cite{chin11} orders with analytically known coefficients,
for example:
$$
{\cal T}_6(\ep)=\frac1{24} {\cal T}_2(\dt)
-\frac{16}{15}{\cal T}_2^2\left(\frac\dt{2}\right)
+\frac{81}{40}{\cal T}_2^3\left(\frac\dt{3}\right),
$$
$$
{\cal T}_8(\ep)=-\frac1{360} {\cal T}_2(\dt)
+\frac{16}{45}{\cal T}_2^2\left(\frac\dt{2}\right)
-\frac{729}{280}{\cal T}_2^3\left(\frac\dt{3}\right)
+\frac{1024}{315}{\cal T}_2^4\left(\frac\dt{4}\right).
$$
Subsitute in ${\cal T}_{2A}$ or ${\cal T}_{2B}$ would then yield RKN
type algorithms.\cite{chin10} Algorithms of up to the 100th-order have been 
demonstrated in Ref.\onlinecite{chin11}.
Since each ${\cal T}_2$ (especially ${\cal T}_{2B}$) only requires a single force-evaluation, 
a $2n$-order algorithm above would require $1+2+\cdots +n=n(n+1)/2$ force evaluations.
Thus at higher orders, RKN's use force evaluations is only quadratic, 
rather than geometric, as in the case of SI.
For order $p=3$ to 6, force subtractions can be combined similarly as in the fourth-order case 
for ${\cal T}_{2A}$ so that the
number of force-evaluation is reduced to $p-1$. Most known RKN algorithms\cite{bat99} 
up to order six can be derived in this manner.\cite{chin10} At order $p=7$, the number
of force-evaluation is 7. At order $p>7$, the number of force-evaluation is greater the $p$.
The distinction between symplectic and RKN algorithms can now be understood as corresponding to a 
single operator product or a sum of operator products.

\section{Forward time-step algorithms}
\la{ftal}

\begin{figure}[t]
\includegraphics[width=0.49\linewidth]{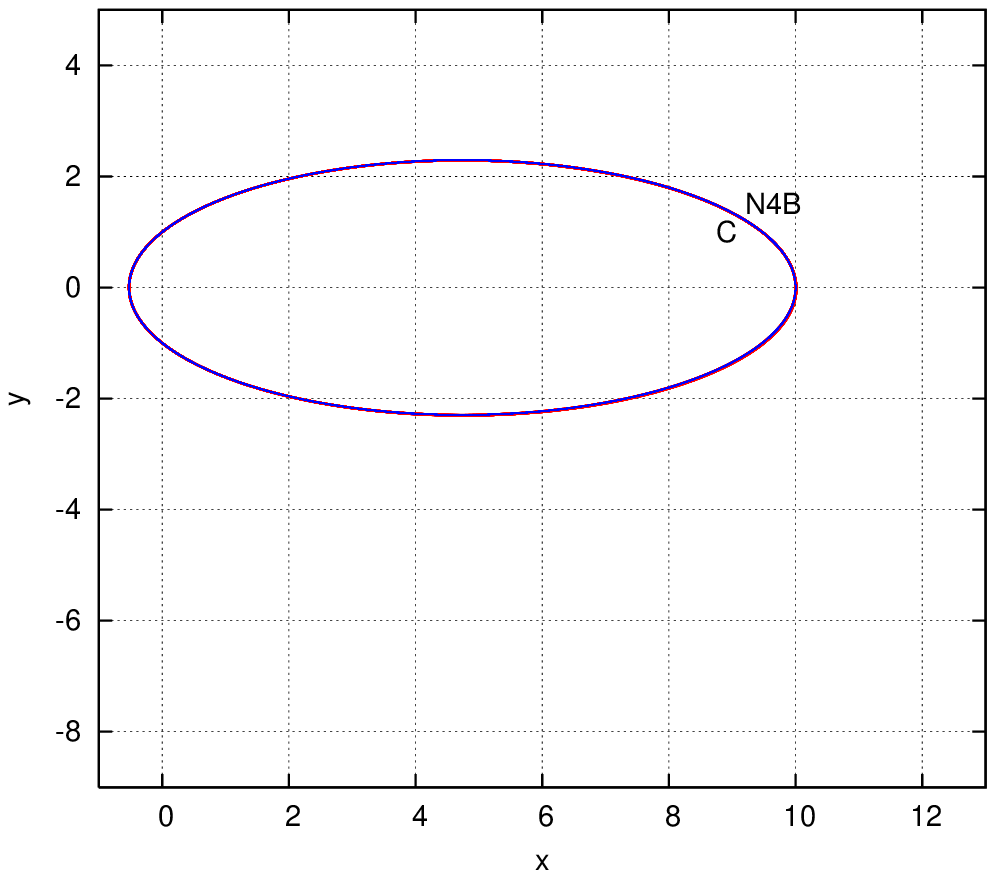}
\includegraphics[width=0.49\linewidth]{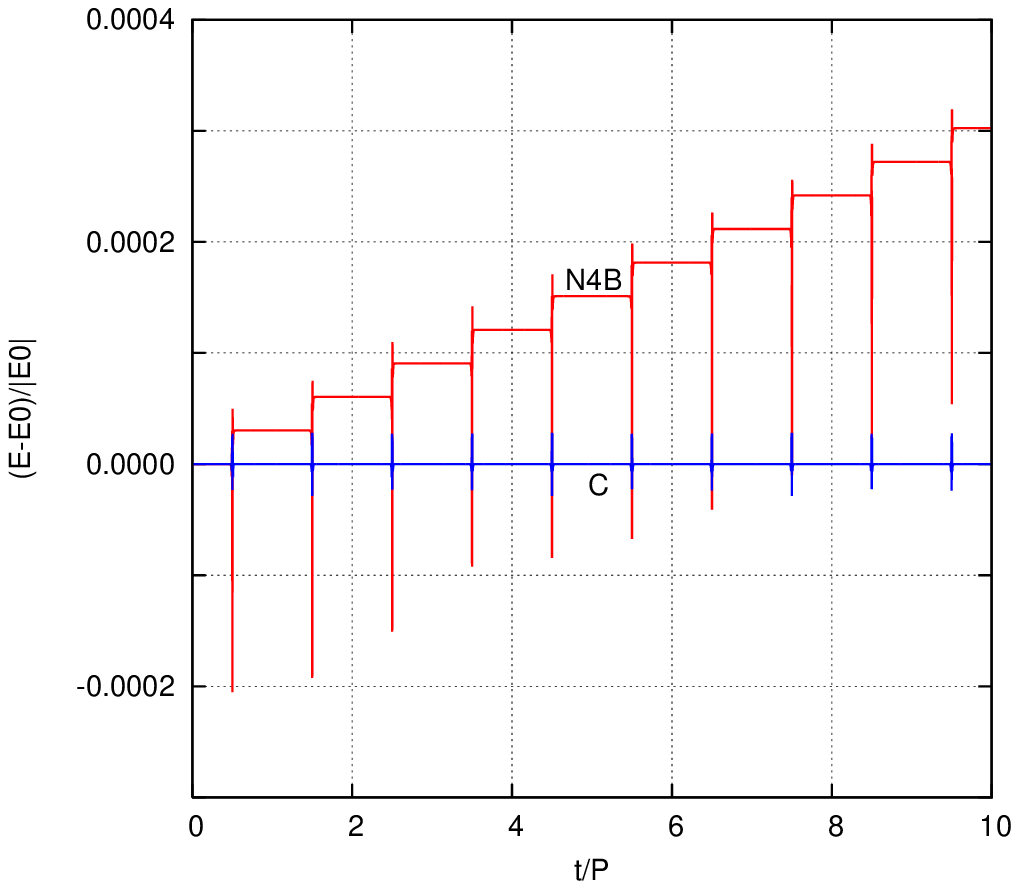} 
\caption{(Color online) Trajectory and energy errors similar to Fig.\ref{eng1} for fourth-order algorithms N4B (red) and C (blue) at $\dt=0.1$. The 40 orbits of both blended into a single purple
orbit.}
\label{eng4f}
\end{figure}

\begin{figure}[t]
\includegraphics[width=0.95\linewidth]{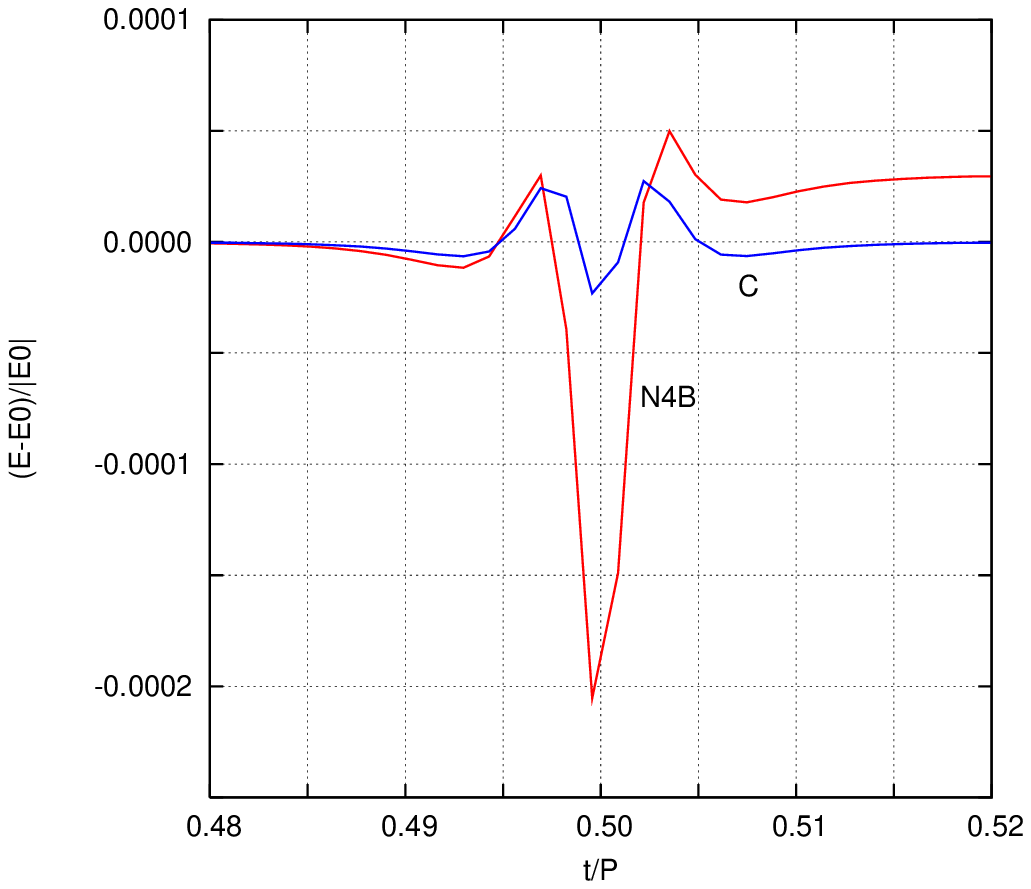}
\caption{(Color online) Energy errors at midperiod for two fourth-order 
algorithms of Fig.\ref{eng4f}.}
\label{eng1p4f}
\end{figure}

Soon after the discovery of the triplet construction (\ref{al4}), it was realized\cite{can91}
that the resulting SI was not that superior to fourth-order RKN type algorithms in the short term, 
as shown in the last section. The problem can be traced to the large negative time-step in (\ref{al4}), which is $\approx -1.7\dt$. It then needs to cancel the two forward time step each at 
$\approx 1.35\dt$, to give back $\dt$. Clearly such a large cancellation is undesirable because
evaluating ${\cal T}_{2}$ at time steps as large as $\approx 1.7\dt$ is less accurate.
To avoid such a cancellation, one might dispense with the triplet construction (\ref{al4}) and
seek to devise SI directly from a single operator product as in (\ref{prod}). 
Unfortunately, even with the general approach of (\ref{prod}), it is not possible to eliminate negative
time steps (and therefore cancellations) beyond second order, 
{\it i.e.}, beyond second order, $\{c_i, d_i\}$ cannot all be positive.
Formally, this result has been proved variously by Sheng,\cite{she89} Suzuki,\cite{suz91} 
Goldman and Kaper,\cite{gol96} 
Blandes and Casas,\cite{blanes05} and Chin\cite{chin06}. However, by repeated application of the BCH
formula, one can show\cite{chin97} that, 

\ba
\e^{\frac13\dt \hT}\e^{\frac3{4}\dt \hV}\e^{\frac23\dt \hT}\e^{\frac14\dt \hV}
&=&\e^{\dt(\hT+\hV)-\frac{1}{48}\dt^3[\hV,[\hT,\hV]]+O(\dt^4)      },\nn\\
\e^{\frac13\dt \hT}\e^{\frac3{4}\dt \hV}\e^{\frac23\dt \hT}\e^{\frac14\dt \hV}
\e^{\frac{1}{48}\dt^3[\hV,[\hT,\hV]]}
&=&\e^{\dt(\hT+\hV)+O(\dt^4)      }.\nn\\
\e^{\frac13\dt \hT}\e^{\frac3{4}\dt \hV}\e^{\frac23\dt \hT}\e^{\frac14\dt \hV
+\frac{1}{48}\dt^3[\hV,[\hT,\hV]]}
&=&\e^{\dt(\hT+\hV)+O(\dt^4)      }.
\la{lal4c}
\ea
Since
$$
[\hV,[\hT,\hV]]=2a_j\frac{\partial a_i}{\partial q_j}\frac{\partial }{\partial v_i}
=\nabla_i (|\ac|^2) \frac{\partial }{\partial v_i},
$$
is just an additional force term, it can be combined together with 
$\hV=a_i (\partial /\partial v_i)$. (Note that $\partial a_i/\partial q_j\propto \partial V/(\partial q_i \partial q_j)=\partial a_j/\partial q_i)$. This additional force is the same as that derived from the 
potential term corresponding to $[\hV,[\hT,\hV]]$ given by (\ref{potvtv}).
All the time-steps on the RHS of (\ref{lal4c}) are then positive.
If one now symmetrized (\ref{lal4c}) to get rid of all the even order $\dt$ error terms, then one arrives at a fourth-order algorithm:
$$
{\cal T}_C(\dt)=
\e^{\frac16\dt \hT}\e^{\frac3{8}\dt \hV}\e^{\frac13\dt \hT}\e^{\frac14\dt (\hV
+\frac{\dt^2}{48}[\hV,[\hT,\hV]])}\e^{\frac13\dt \hT}\e^{\frac3{8}\dt \hV}\e^{\frac16\dt \hT}.
$$
We leave it as an exercise for the readers to write out the algorithmic form of this {\it forward time-step} algorithm C of Ref.\onlinecite{chin97}.

In Fig.\ref{eng4f}, we compare the working of algorithm C with N4B. At $\dt=0.1$, both algorithms are able to maintain a single orbit after 40 periods. These are the two best fourth-order algorithms when tested on the Kepler problem.\cite{chin10} Again, one sees that N4B's energy error is increasing with each orbit while that of C remained a periodic spike at midperiod. In Fig.\ref{eng1p4f}, N4B's maximum energy error is befittingly $O(\dt^4)$, but that of C's is an order of magnitude smaller.

Both N4B and C are new algorithms, not known from classical
Runge-Kutta type analysis. Currently, it is not possible to extend forward time-step
algorithms, like C, beyond the fourth-order.\cite{nosix} A sixth-order scheme
would have required an additional mixed product term of momenta and coordinates,
corresponding to a {\it non-separable} Hamiltonian, which is difficult to solve.

\section{Concluding summary}
\la{sum}

     In this work, we have outlined the ``modern synthesis" of numerical algorithms for solving 
classical dynamics. The key findings are: 1) There is no arbitrariness in the choice of elementary algorithms.
There are only two standard, canonical, and time-reversible second-order algorithms 
(\ref{al2a}) and (\ref{al2b}), out of which nearly all higher-order symplectic {\it and} RKN
algorithms can be derived. 2) The stability of symplectic integrators 
can be simply understood as due to their {\it sequential} updating of dynamcial variables, 
which is a fundamental but under-appreciated property of canonical transformations. 3)
The difference between symplectic and non-symplectic algorithms is most easily understood
in terms of exponentials of Lie operator. The former is due to a single product 
of exponentials while the latter is due to a sum of products. From this prospective, it is also easy 
to see why the former obeys Liouville's theorem, while the latter does not. Because of this,
for periodic motion, the energy errors of SI remain periodic while those of RKN algorithms
increase monotonically.
4) In the last Section, we have introduced and compared two state-of-the-art, yet to be widely
known fourth-order algorithms.

All algorithms derived here can be easily generalized to solve dynamics due to a time-dependent force.\cite{chin11} In physics, the only important velocity-dependent force is the Lorentz force acting on
a charged particle in a magnetic field. In this case, there are also similar sequential algorithms
\cite{chin08} which can conserve the energy exactly.

All algorithms described here in solving for the evolution operator $\e^{\dt(\hT+\hV)}$ can be applied to
any other operators $\hT$ and $\hV$ if $\dt$ is allowed to be both positive and negative, such as the
time-dependent Schr\"odinger equation. If $\dt$ must be positive, as in solving a diffusion-related equation,
then only forward time-step algorithms can be used, as in the case of the imaginary-time 
Schr\"odinger equation. 


\end{document}